\documentclass[acmsmall, anonymous=false]{acmart}

\usepackage[export]{adjustbox}
\usepackage{wrapfig}
\usepackage{graphicx}
\usepackage{algpseudocode} 
\usepackage[mathscr]{eucal}
\usepackage{amsmath}
\usepackage{mathtools}
\usepackage{multicol}
\usepackage{algorithm} 
\usepackage{bm}
\usepackage{hyperref}
\usepackage{booktabs}
\usepackage[applemac]{inputenc}
\usepackage{xparse}
\usepackage{xifthen}
\usepackage{mathtools}
\usepackage{multirow}
\usepackage{textcomp}
\usepackage{todonotes}
\usepackage{pifont}
\usepackage{soul}
\usepackage{color}
\usepackage{upgreek}
\usepackage{color,soul}
\newcommand{\cmark}{\ding{51}}%
\renewcommand{\textsl}{\textit}
\newcommand{\comm}[1]{}

\NewDocumentCommand{\vect}{ O{} O{} m }{\mathbf{#3}\ifthenelse{\isempty{#1}}{}{^{(#1)}}\ifthenelse{\isempty{#2}}{}{_{#2}}}

\NewDocumentCommand{\mat}{ O{} O{} m }{\mathbf{#3}\ifthenelse{\isempty{#1}}{}{^{(#1)}}\ifthenelse{\isempty{#2}}{}{_{#2}}}

\NewDocumentCommand{\ten}{ O{} O{} m }{\pmb{\mathscr{#3}}\ifthenelse{\isempty{#1}}{}{^{(#1)}}\ifthenelse{\isempty{#2}}{}{_{#2}}}
\AtBeginDocument{%
  \providecommand\BibTeX{{%
    \normalfont B\kern-0.5em{\scshape i\kern-0.25em b}\kern-0.8em\TeX}}}

\setcopyright{acmcopyright}
\copyrightyear{2021}
\acmYear{2021}
\acmDOI{10.1145/xxxxxxx.xxxxxxx}

\acmConference[TOPS]{TOPS: ACM Transactions on Privacy and Security}{December 24, 2021}{Woodstock, NY}
\acmBooktitle{TOPS: ACM Transactions on Privacy and Security,
  December 24, 2021, Woodstock, NY}
\acmPrice{15.00}
\acmISBN{978-1-4503-XXXX-X/18/06}

\begin{document}

\title[Semi-supervised Classification of Malware Families]{Semi-supervised Classification of Malware Families \\Under Extreme Class Imbalance via\\Hierarchical Non-Negative Matrix Factorization\\with Automatic Model Selection}

\author{Maksim E. Eren}
\affiliation{%
  \institution{Advanced Research in Cyber Systems, LANL}
  \country{USA}
}
\email{maksim@lanl.gov}

\author{Manish Bhattarai}
\affiliation{%
  \institution{Theoretical Division, LANL}
  \country{USA}
}
\email{ceodspspectrum@lanl.gov}

\author{Robert J. Joyce}
\affiliation{%
  \institution{Machine Learning Research Group, Booz Allen Hamilton}
  \country{USA}
}
\email{ceodspspectrum@lanl.gov}

\author{Edward Raff}
\affiliation{%
  \institution{Machine Learning Research Group, Booz Allen Hamilton}
  \country{USA}
}
\email{raff\_edward@bah.com}

\author{Charles Nicholas}
\affiliation{%
  \institution{Department of Computer Science and Electrical Engineering, UMBC}
  \country{USA}
}
\email{nicholas@umbc.edu}

\author{Boian S. Alexandrov}
\affiliation{%
  \institution{Theoretical Division, LANL}
  \country{USA}
}
\email{boian@lanl.gov}

\renewcommand{\shortauthors}{Eren et al.}

\begin{abstract}
Identification of the family to which a malware specimen belongs is essential in understanding the behavior of the malware and developing mitigation strategies. Solutions proposed by prior work, however, are often not practicable due to the lack of realistic evaluation factors. These factors include learning under class imbalance, the ability to identify new malware, and the cost of production-quality labeled data. In practice, deployed models face prominent, rare, and new malware families. At the same time, obtaining a large quantity of up-to-date labeled malware for training a model can be expensive. In this paper, we address these problems and propose a novel hierarchical semi-supervised algorithm, which we call the \textit{HNMFk Classifier}, that can be used in the early stages of the malware family labeling process. Our method is based on non-negative matrix factorization with automatic model selection, that is, with an estimation of the number of clusters.  With \textit{HNMFk Classifier}, we exploit the hierarchical structure of the malware data together with a semi-supervised setup, which enables us to classify malware families under conditions of extreme class imbalance. Our solution can perform abstaining predictions, or rejection option, which yields promising results in the identification of novel malware families and helps with maintaining the performance of the model when a low quantity of labeled data is used. We perform bulk classification of nearly 2,900 both rare and prominent malware families, through static analysis, using nearly 388,000 samples from the EMBER-2018 corpus. In our experiments, we surpass both supervised and semi-supervised baseline models with an F1 score of 0.80.
\end{abstract}

\ccsdesc[500]{Security and privacy~Malware and its mitigation}
\ccsdesc[500]{Computing methodologies~Semi-supervised learning settings}
\ccsdesc[500]{Computing methodologies~Non-negative matrix factorization}
\ccsdesc[500]{Computing methodologies~Topic modeling}

\keywords{malware, malware families, non-negative matrix factorization, semi-supervised, hierarchical, model selection, class imbalance, abstaining prediction, reject-option}

\maketitle

\section{Introduction}

The objective of malware detection is to identify a given file as benign or malicious, typically by using its run-time behavior (dynamic malware analysis) and/or static information (static malware analysis). In contrast to malware detection, malware family classification  assumes that any given sample is already known to be malicious, and we want to know which family it belongs to \cite{Raff2020ASO}. New malware samples are created regularly by threat actors by various techniques, which create new versions of already existing malware specimens with identical functionality \cite{Raff2020ASO}. Malware analysts regularly go through large quantities of malware samples to understand if a new specimen in fact belongs to a previously known malware family. Classifying a new malware sample into a family can reduce the number of files analysts need to examine, and aid in understanding the behavior of the malware; this is in turn helpful for estimating the severity of the threat, developing mitigation strategies, and building datasets \cite{Raff2020ASO}. The tools that can aid in malware detection and classification are especially significant now as recent reports point out that malware is one of the most frequent and costly cyber threats \cite{Ponemon2019}. 

Approximately half a million new malware specimens are reported daily, which drives the increased utilization of Machine learning (ML) based automated security systems to combat malware \cite{avtest_2021, WinIntel, VirusTotalSangfor, VirusTotalBitdefender, kaspersky_ml}. However, the adoption of ML-based solutions against malware threats has been relatively slow despite the cost savings \cite{ibm2021}. Shortcomings in the existing solutions are perhaps contributing to this problem. The majority of prior research for malware family classification, over the past two decades, has not sufficiently accounted for core evaluation criteria in their work including learning under class imbalance, ability to identify new malware, and the cost of production-quality labeled data \cite{nguyen2021leveraging, Raff2020ASO}. For example, the majority of ML solutions for malware family classification are unrealistically limited to identifying the top most populous families. This results in reports of excellent performance on evaluation metrics that do not generalize to the real world, limited as they have been to the analysis of \textit{``easy''} malware. At the same time, semi-supervised learning in the malware classification field has not been widely explored despite its potential benefits \cite{Raff2020ASO}. With the ever-growing quantity of malware, attacks, and their complexities there is an urgent need to improve existing solutions and their operational architectures to drive the increased adaption of ML-based solutions.

In this work, we introduce a novel semi-supervised algorithm, named Hierarchical Non-Negative Matrix Factorization with automatic model selection Classifier (or \textit{HNMFk Classifier}). The HNMFk Classifier classifies Windows Portable Executable (PE) format malware specimens (e.g. from the EMBER-2018 dataset) into families using static malware analysis-based features \cite{2018arXiv180404637A}. Our method performs bulk classification where the known samples are used as a reference against the unknown specimens when performing hierarchical clustering, resulting in a model with only an inference process (i.e. no training). Therefore, in comparison to the traditional ML models which have separate training (slow) and prediction (fast) steps, our solution can be used outside the real-time environments, such as early stages in the labeling process of the malware. \textit{HNMFk Classifier} performs hierarchical clustering using Non-Negative Matrix Factorization (NMF) with automatic model selection (\textit{NMFk}) \cite{alexandrov2013deciphering,alexandrov2013signatures, alexandrov2020patent, SmartTensors, nebgen2021neural,eren2023malwaredna,eren2022senmfk,eren2022one}, which helps us determine whatever hierarchical structure exists among the malware specimens. 
With a semi-supervised setting, we obtain the ability to perform abstaining predictions (i.e. predicting "I do not know"), in addition to answering either "Yes" or "No". Specifically, our model incorporates a reject option, where it abstains from making a prediction (reject). Abstaining predictions aid in detecting novel malware, reduce the need to include all malware families during factorization to achieve good generalizability to new malware, and help our model to maintain its performance with a low quantity of labeled data. 

In our experiments, we first use a small subset of the dataset (in a setup that does not reflect the real world) to understand the effects of using different hyper-parameters in our model, conduct ablation studies, and to observe the performance of our model with a decreasing quantity of labeled data. During our (more realistic) larger scale experiments, we use 2,898 classes of malware families (numbering more than 388,000 samples) with extreme class imbalance, and while including novel unknown malware samples during classification. Our method surpasses the supervised baseline models \textit{XGBoost} and \textit{LightGBM} \cite{xgboost, 10.5555/3294996.3295074}. We further extend these baselines with the \textit{SelfTrain} algorithm to create strong semi-supervised models, which our approach still outperforms \cite{10.3115/981658.981684}. We also achieve better classification results compared to our Multilayer Perceptron (MLP) baseline \cite{haykin1994neural}. To the best of our knowledge, we are the first to perform malware family classification over the EMBER-2018 corpus under realistic conditions such as the inclusion of the rare and novel families during our experiments, and our target number of family classes is around 29 times more than the previous work with the largest number of classes \cite{huang2016mtnet}. Our contributions include:
\begin{itemize}
    \item Introducing a novel semi-supervised hierarchical bulk classifier, the \textit{HNMFk Classifier}, that can assist analysts early in the malware family labeling process.
    \item Identifying Windows malware families using static malware analysis-based features, specifically using malware meta-data and PE header features, under extreme class imbalance conditions.
    \item Utilizing abstaining prediction to enable our model to help identification of novel malware families, and maintain its accuracy as the amount of labeled data decreases.
    \item Achieving higher F1 scores compared to the baseline supervised and semi-supervised learners which prior work used to report their benchmarks when classifying malware families in the EMBER-2018 corpus.
\end{itemize}

The remainder of the paper is organized as follows: we provide a summary of related work in Section \ref{sec:relevant_work}. Section \ref{sec:methods} includes a description of NMF (Section \ref{sec:nmf}), automatic model selection with \textit{NMFk} (Section \ref{sec:nmfk}), and hierarchical NMF (Section \ref{sec:hierarhicalnmf}). We then introduce our \textit{HNMFk Classifier} in Section \ref{sec:hNMFkclassifier}. Section \ref{sec:dataset} describes the dataset and the features used in our experiments, pre-processing of the features, and the preparation of the experiments.  Section \ref{sec:results} showcases our experimental results including a performance analysis over a subset of the data and how our model preserves accuracy under the low quantity of labeled data in Section \ref{sec:performance_analysis}. Our results when classifying the malware families under realistic conditions, and comparison to the baseline models, are shown in Section \ref{sec:large_scale_classification}. We justify the parts of our methodology with ablation studies in Section \ref{sec:ablation_studies}. Before concluding, we list potential areas of future work to explore in Section \ref{sec:future_work}.

\section{Related Work}
\label{sec:relevant_work}

Malware classification is a challenging task, and the quantity and complexity of malware continues to increase rapidly. This makes the ML-based malware classification an important field of study. Raff et al. surveys over 200 research articles on ML-based malware analysis \cite{Raff2020ASO}. This survey of the field emphasizes that the standard ML model evaluation technique, where the dataset containing malware families are divided into training and test sets, is flawed when it comes to the malware family classification problem in the real-life case, since previously unseen malware families will continue to appear. To this end, they recommend that the ability to perform abstaining prediction can assist analysts in identifying novel malware. However, prior work has not widely studied this open problem area for malware classification. In our experiments, we evaluate the performance of our solution by including a set of malware families that were not present in the known set. Additionally, Raff et al. discuss the challenges in malware data gathering, and the expensive and time-consuming process of file labeling. Their survey found that semi-supervised solutions are not yet fully explored, although they can help when faced with only a small quantity of labeled data. In support of this finding, we show that our solution continues to maintain its performance with the decreasing amount of labeled malware in our small-scale experiment in Section \ref{sec:performance_analysis}. Finally, Raff et al. also point out the relatively small amount of prior work on the problem of class imbalance. This issue was also emphasized by Nguyen et al. since much prior work has unrealistically evaluated their solutions over the top most populous malware families \cite{nguyen2021leveraging}. Our study addresses this problem by including both the rare and prominent classes of malware families during the large-scale experiment in Section \ref{sec:large_scale_classification}. 

Several previous works have looked at malware family classification, however, they tend to use only the most common malware families, did not consider novel malware families, or used manually balanced datasets when reporting their results \cite{8858297, loi2021towards, 8681127, 8322598, 7440587, ahmadi2016novel, 8293854, bak2020clustering, jiang2019android}. 
In contrast, when comparing to the baseline models, we report our results when classifying specimens belonging to the whole ensemble of malware families present in the EMBER-2018 dataset with an imbalanced setup which also includes novel unknown specimens (Section \ref{sec:large_scale_classification}). This setup allows our results to be more like what malware analysts would see in the real world. Several prior works also considered class imbalance, however, they still targeted a small number of top malware families, and rare specimens are mapped to a single \textit{"others"} class \citep{loi2021towards, MOHAISEN2015251}. To the best of our knowledge, the most realistic and the largest malware family classification work was done by Huang et al. \cite{huang2016mtnet}, which targeted 100 classes where two of the classes include the one for benign samples and another for the rare specimens. This type of setup, although it considers class imbalance, limits the classification capabilities to only a handful of malware families. In contrast, we do not map the rare specimens into a single class, but rather recognize all 2,898 malware families as individual classes. Furthermore, supervised methods used in prior work often poorly generalize to rare specimens as also pointed out by Loi et al \cite{loi2021towards}. Loi et al. reports that their false positives are heavily represented by the families collected within the \textit{"others"} class due to the supervised method's inability to learn the patterns of these families from a rare number of specimens. We use a semi-supervised approach, which has an added benefit of improved generalizability and ability to work with a low quantity of labeled data compared to the supervised models. A number of these prior works did consider benign-ware as a class in their analysis \cite{8681127, loi2021towards, MOHAISEN2015251}, but we assume the samples are already known to be malware and perform only malware family classification. We summarize the mentioned prior work and show how they compare to our research in Table \ref{table:prior_work}. 

\begin{table*}[htb]
\vspace{-1.1em}
\caption{The comparison of prior and our work in dataset size, number of classes, consideration of imbalanced data and novel malware families, and the method used. \textit{Custom} refers to the proprietary datasets, or the custom build datasets by the authors.}
 \label{table:prior_work} 
\resizebox{\columnwidth}{!}{
\begin{tabular}{c|c|c|c|c|c|c}
\hline
\textbf{Reference}            & \textbf{Dataset(s)}  & \textbf{Dataset Size}        & \textbf{Num. Classes}  & \textbf{Imbalanced Data} & \textbf{Novel Malware} & \textbf{Method}\\ \hline

\textbf{Ours}           & EMBER-2018 \cite{2018arXiv180404637A}                      & 388k                         & 2,898                  & \cmark              & \cmark                 & Semi-supervised \\
\cite{huang2016mtnet}   & \textit{Custom}                                            & 6.5m                         & 100                    & \cmark              & ---                 & Supervised      \\
\cite{jiang2019android} & Drebin \cite{arp2014drebin}                                & 5k                           & 40                     & ---              & ---                 & Supervised      \\
\cite{8681127}          & Malimg \cite{10.1145/2016904.2016908} \& \textit{Custom}   & 9k \& 10k                    & 25 \& 10               & ---              & ---                 & Supervised      \\
\cite{bak2020clustering}& \textit{Custom}                                            & 10k                          & 14                     & \cmark              & ---                 & Unsupervised     \\
\cite{8858297}          & EMBER-2018 \cite{2018arXiv180404637A}                      & 750k                         & 21                     & ---              & ---                 & Supervised          \\
\cite{loi2021towards}   & EMBER-2017 \cite{2018arXiv180404637A}                      & 500k                         & 21                     & \cmark              & ---                 & Supervised         \\
\cite{8322598}          & VirusShare \cite{virusShare}                               & 2.7k                         & 12                     & ---              & ---                 & Supervised    \\
\cite{ahmadi2016novel}  & Malimg \cite{10.1145/2016904.2016908}                      & 21k                          & 9                      & ---              & ---                 & Supervised          \\
\cite{MOHAISEN2015251}  & \textit{Custom}                                            & 115k                         & 8                      & \cmark              & \cmark                 & Supervised              \\
\cite{7440587}          & \textit{Custom}                                            & 31k                          & 5                      & ---              & ---                 & Supervised    \\
 
\hline
\end{tabular}
}
\end{table*}

Non-negative Matrix Factorization, or NMF, has also been applied to the malware/benign-ware classification problem. Ling et al. derive similarity scores of structural patterns extracted with NMF to detect metamorphic malware (malware with the capability to modify its code during run-time) using static analysis features \cite{ling2019nonnegative}. In their experiments, they choose a fixed number of components for NMF where the number of components $k$ is selected as $k(n + m) < nm$. A single application of NMF misses the patterns hidden in malware sub-groups, and using a fixed number of components can result in missing important information (under-fitting) or including noise (over-fitting) in the results. Unlike Ling et al., we perform malware family classification by applying hierarchical NMF to discover the sub-groups and utilize \textit{NMFk} as a heuristic to determine the number of components or clusters. Prior work outside the malware analysis field has demonstrated that hierarchical NMF can be used to achieve good clustering of the data \cite{trigeorgis2014deep, gillis2014hierarchical}. Gillis et al. show that using rank-two factorization at each step (i.e. split the data into two at each stage, $k=2$) yields good clustering results when applied with hierarchical NMF \cite{gillis2014hierarchical}. We use hierarchical rank-two NMF in our ablation studies and show that estimating the number of components via \textit{NMFk} produces better classification results, although extracting two clusters at each factorization does yield good classification results that surpass our baseline models.

\section{Methods}
\label{sec:methods}

Our work draws on prior advances in standard and hierarchical NMF methods, and automatic model selection. In this section, we give a brief summary for each of these methods, then we introduce our \textit{HNMFk} Classifier. The summary of the notations used throughout the paper is provided in Table \ref{table:notations}.

\begin{table*}[htb]
\caption{Summary of the notation styles used in the paper.}
 \label{table:notations}
\begin{tabular}{c|l}
\hline
\textbf{Notation}  & \textbf{Description}  \\ \hline
$x$                &  Scalar\\
$\vect{x}$         &  Vector\\
$\mat{X}$          &  Matrix\\
$\ten{X}$          &  Tensor\\
$\vect{x}_i$       &  $i$th element in the vector\\
$\mat{X}_{ij}$    & Entry located on row $i$ and column $j$\\
$\mat{X}_{i:}$   & $i$th row \\
$\mat{X}_{:j}$   & $j$th column \\
$\ten{X}_{::i}$   & $i$th slice along the third dimension\\
$\ten{X}_{:::i}$   & $i$th slice along the fourth dimension\\
$\ten{X}^{\textit{name}}$  & Superscript \textit{name} used as an identifier \\
$*$               & Dot product
\\\hline
\end{tabular}
\end{table*}

\subsection{Non-negative Matrix Factorization (NMF)}
\label{sec:nmf}

NMF is an unsupervised learning method based on a low-rank matrix approximation. NMF represents an observed non-negative matrix, $\mat{X}\in \mathbb{R}_{+}^{n \times m}$, as a product of two (unknown) non-negative matrices, $\mat{W} \in \mathbb{R}_{+}^{n\times k}$, and $\mat{H} \in \mathbb{R}_{+}^{k \times m}$, where usually $k\ll m, n$. Here, $n$ is the number of samples, and $m$ is the number of features. This approximation is performed via non-convex minimization with a given distance, $||...||_{dist}$, constrained by the non-negativity of $\mat{W}$ and $\mat{H}$: min$||\mat[][ij]{X}-\sum^{k} _{s=1}\mat[][is]{W} \mat[][sj]{H}||_{dist}$. NMF relies on a generative statistical model predetermined by the choice of the distance $|| . . . ||_{dist}$. For example, if the Frobenius norm is chosen as a distance, NMF can be treated as a Gaussian mixture model \cite{fevotte2009nonnegative}. If KL-divergence is chosen, we have a generative Poisson model \cite{canny2004gap}, equivalent to latent Dirichlet allocation under uniform Dirichlet prior \cite{de2016equivalence}. In both cases, the number of latent features of the superimposed components is equal to the size of the small dimension $k$, and NMF minimization is equivalent to the expectation-minimization (EM) algorithm. In this probabilistic interpretation of NMF, the observables are the rows of $\mat{X}$ generated by latent variables, the rows of the matrix $\mat{W}$, with weights (the basis patterns), represented by the columns of matrix $\mat{H}$. Thus, each row $\mat[][i:]{X}$ of $\mat{X}$ is generated from a probability distribution with mean $\mat[][i:]{X} =\sum^{k}_{ s=1}\mat[][is]{H} \mat[][s:]{W}$.

\subsection{Automatic Model Selection: NMFk}
\label{sec:nmfk}
The NMF minimization requires prior knowledge of the latent dimensionality, $k$ (the number of latent features), which is usually unavailable. It is known that choosing too small a value of $k$ leads to a poor approximation of the observables in $\mat{X}$ (\textit{under-fitting}), while if $k$ is chosen to be too large, the extracted features are not easily explainable because they also fit the noise in the data (\textit{over-fitting}).  In other words, choosing $k$ is equivalent to estimating the number of parameters of the model, which is a difficult and a well-known problem. 

In general, the existing partial solutions of this problem are heuristic. Among these solutions is Automatic Relevance Determination (ARD) \cite{mackay1994bayesian} which was first modified for Principal Component Analysis \cite{bishop1999bayesian}, and then for NMF \cite{morup2009tuning,tan2012automatic}.
Another approach is based on an assumed stability of the NMF solution, and was proposed to identify the number of stable clusters in the observational matrix $\mat{X}$ \cite{brunet2004metagenes}. A recent model selection technique, called \textit{NMFk} \cite{alexandrov2020patent}, has been successfully used to decompose the largest collection of human cancer genomes \cite{alexandrov2013signatures}. 
\textit{NMFk} integrates classical NMF-minimization with custom clustering and Silhouette statistics \cite{ROUSSEEUW198753}, and combines the accuracy of the minimization and robustness/stability of the NMF solutions, when a bootstrap procedure (i.e., generation of a random ensamble of slightly perturbed input matrices) is applied  to estimate the number of latent features, see for example, \cite{alexandrov2013deciphering}. Recently, \textit{NMFk} was applied to a large number of synthetic datasets with a predetermined number of latent features, and it was demonstrated its superior performance of correctly estimating $k$ in comparison to the other known heuristics \cite{nebgen2021neural}. The  superior performance of NMFk method as a model selection was also demonstrated in identifying mutational genome signatures in a large set of cancer genomes, both in practice \cite{alexandrov2020repertoire} and in large set of synthetic cancer genomes with predetermined number of latent features \cite{islam2022uncovering}. In addition, it was shown that \textit{NMFk} performs better than spherical k-means and other methods for topic extraction \cite{vangara2020semantic}. Our numerical experiments here demonstrate that NMFk performs better than the predetermined k=2 case. Therefore, we use \textit{NMFk} as the core factorization method with automatic model selection needed to extract the right clusters of malware, after the NMF dimension reduction.
In this work, we are making extensive use of \textit{NMFk}, and for completeness we provide the pseudocode for it in Algorithm 1 and a description of it, as follows:
\begin{algorithm}[htb]
    \caption{NMFk($\mat{X}$, $k^{min}$, $k^{max}$, $M$, $Sill\_{thr}=0.8$)} 
	\begin{algorithmic}[1]
	    \Require: $\mat{X} \in \mathbb{R}_{+}^{n \times m}$ , $k^{min}$, $k^{max}$ , $r$ \label{alg:nmfk}
		\For {$k$ in $k^{min}$ to $k^{max}$} \Comment{Start and end process for NMFk}
			\For{$q$ in 1 to $M$}        \Comment{Num. of Perturbations on each k}
				\State $\ten{X}_{::q}$ = Perturb($\mat{X}$) \Comment{Resampling  $\mat{X}$ to create a random ensemble}
				\State $\ten{W}_{::kq}$,$\ten{H}_{::kq}$ = NMF($\ten{X}_{::q}$,k)
				
			\EndFor
			\State  $\ten{W}^{all}$=[$\ten{W}_{::k1}$,\ldots,$\ten{W}_{::kM}$] and $\ten{H}^{all}$=[$\ten{H}_{::k1}$,\ldots,$\ten{H}_{::kM}$]
			\State $\ten{\hat{W}}$, $\ten{\hat{H}}$ = customCluster( $\ten{W}^{all}$,$\ten{H}^{all}$)
			\State $\ten{\widetilde{W}}_{::k}$ = medians( $\ten{\hat{W}}$) 
			\State $\ten{H}^{reg}_{::k}$  = NNLS($\mat{X}$,$\widetilde{\mat{W}}_{::k}$) \Comment{Column-wise regression of $\mat{H}$ with $\widetilde{\mat{W}}$ and column of $\mat{X}$}
			\State $\vect{s}_{k}$ = clusterStability($\ten{\hat{W}}$)
			\State $\vect{err_{k}}$ = reconstructErr($\mat{X}$,$\widetilde{\mat{W}}_{::k}$ , $\mat{H}^{reg}_{::k}$ ) \Comment{Column-wise reconstruction error for L-statistics}
	\EndFor

	\State $\vect{err}^{all}$=[$\vect{err}_{k^{min}}$,\ldots,$\vect{err}_{k^{max}}$] 
	\State $k^{opt}$ = PvalueAnalysis($\vect{err}^{all}$ ,$k^{min}$,$k^{max}$,$\vect{s}_{k}$,$Sill\_{thr}$) \Comment{Predicted k value using Wilcoxon}		
    \State \textbf{return} $\ten{\widetilde{W}}_{::k^{opt}}$, $\ten{H}^{reg}_{::k^{opt}}$, $k^{opt}$
	\end{algorithmic} 

\textbf{Ensure:} $k =k^{opt}$,$\ten{\widetilde{W}}_{::k^{opt}}$ $\in \mathbb{R}_{+}^{n \times k}$ ,$\ten{H}^{reg}_{::k^{opt}}$ $\in \mathbb{R}_{+}^{k \times m}$ , $\mat{X}$ = $\ten{\widetilde{W}}_{::k^{opt}}$  $\ten{H}^{reg}_{::k^{opt}}$	
\end{algorithm}
\begin{enumerate}
    \item \textit{Resampling}: Based on the observable matrix, $\mat{X}$, \textit{NMFk} creates an ensemble of $M$ random matrices, $[\ten{X}_{::q}]_{q=1,...,M}$, with means equal to the original matrix $\mat{X}$. Each one of these random matrices $\ten{X}_{::q}$ is generated by perturbing the elements of $\mat{X}$ by a small uniform noise, such that: $\ten{X}_{ijq} = \mat{X}_{ij} +\delta$, for each $q=1,...,M$, where $\delta$ is the small error.
    \item \textit{NMF minimization}: We use the Frobenius norm-based multiplicative updates (MU) algorithm \cite{lee1999learning} to explore different numbers of latent features, $k$, in an interval $[k^{min},k^{max}]$, for each one of the generated $M$ random matrices.
    \item \textit{Custom clustering:} For each $k\in [k^{min},k^{max}]$, NMF minimizations of the $M$ random matrices, $[\ten{X}_{::q}]_{q=1,...,M}$, results in $M$ pairs $[\ten{W}_{::kq}; \ten{H}_{::kq}]_{q=1,...,M}$. Further, \textit{NMFk} clusters the set of the $M*k$ latent features, the columns of $\ten{W}_{::kq}$. The \textit{NMFk} custom clustering is similar to k-means, but it holds in each one of the  clusters exactly one column from each of the $M$ NMF solutions. This constraint is needed since each NMF minimization gives exactly one solution $\ten{W}_{::kq}$ with the same number of columns, $k$. In the clustering, the similarity between the columns is measured by the cosine similarity metric.
    \item \textit{Robust $\mat{W}$ and $\mat{H}$ for each $k$:} The medians of the clusters,
    $\ten{\widetilde{W}}_{::k}$,  are the robust solution for each explored $k$. The corresponding mixing coefficients $\ten{H}^{reg}_{::k}$ are calculated by regression of $\mat{X}$ on  $\ten{\widetilde{W}}_{::k}$.
    \item \textit{Cluster stability via Silhouette statistics:} \textit{NMFk} explores the stability of the obtained clusters, for each $k$, by calculating their Silhouettes \cite{ROUSSEEUW198753}. Silhouette statistics quantify the cohesion and separability of the clusters. The Silhouette values range between $[-1, 1]$, where $-1$ means an unstable cluster, while $+1$ means perfect stability.
    
    \item \textit{Reconstruction error:} Another metric \textit{NMFk} uses is the relative reconstruction error, $R = ||\mat{X} - \ten{X}_{::k}^{rec}||/||\mat{X}||$, where $\ten{X}_{::k}^{rec} = \ten{\widetilde{W}}_{::k}*\ten{H}^{reg}_{::k}$, which measures the accuracy of the reproduction of initial data by a given solution and the number of latent features $k$. 
    \item \textit{L-statistics:} \textit{NMFk} uses L-statistics \cite{vangara2021finding} to automatically estimate the number of latent features. To calculate L-statistics for each $k$, \textit{NMFk} records the distributions of the column reconstruction errors, $\vect{e}_i = \Vert \mat{X}_{:j} - \ten{X}^{rec}_{:jk}\Vert/\Vert\mat{X}_{:j}\Vert$; $j=1,...,m$. L-statistics compares the distributions of column errors for different $k$ by a two-sided Wilcoxon rank-sum test \cite{Haynes2013}, which evaluates whether two samples are taken from the same population. 
    \item \textit{NMFk final solution:} The number of latent features, $k^{opt}$, is determined as the maximum number of stable clusters corresponding to a good accuracy of the reconstruction. The Wilcoxon rank-sum test determines the p-value of the given $k^{opt}$. \textit{NMFk} is "looking" for a distribution of the column errors such that the next distributions (each one with bigger $k$) are statistically the same, and the model is fitting the noise. The L-statistics used in conjunction with the condition that the \emph{minimum} Silhouette be greater than $0.80$. The threshold of $0.80$ is selected to place the predicted $k^{opt}$ prior to a steep decline in the \emph{minimum} Silhouette. The corresponding $\ten{\widetilde{W}}_{::k^{opt}}$ and $\ten{H}^{reg}_{::k^{opt}}$ are the robust solutions for the low-rank factor matrices.
\end{enumerate}

\begin{figure}[htb]
    \centering
    \includegraphics[width=0.8\textwidth]{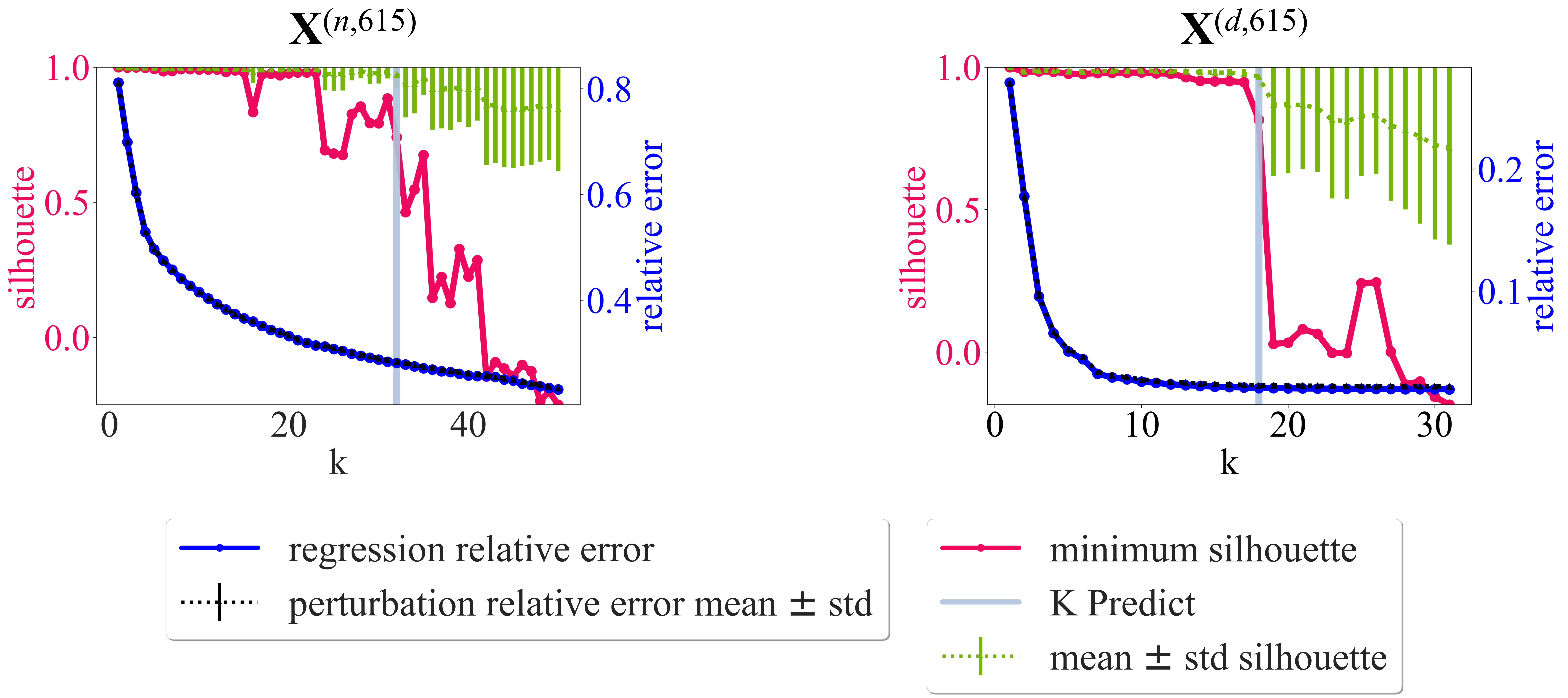}
    \caption{Sample Silhouette and relative error graphs obtained from NMFk is shown for the matrices $\mat{X}^{(n,615)}$ and $\mat{X}^{(d,615)}$ which are formed using 1,000 malware specimens from 10 families. $\mat{X}^{(d,615)}$ consist of samples extracted from a single cluster after the the first NMFk procedure on $\mat{X}^{(n,615)}$.}
    \label{fig:nmfk_sil_plots}
\vspace{-1.3em}
\end{figure}

We provide a sample Silhouette score and relative error plot produced by \textit{NMFk} for two factorizations, to demonstrate the selection of $k$, in Figure \ref{fig:nmfk_sil_plots}. The presented \textit{NMFk} framework estimates the latent feature count based on two criteria, namely a high minimum Silhouette score, and a low relative reconstruction error, which corresponds to a stable NMF solution. The number of features with lower minimum Silhouette scores correspond to overlapping clusters or scattered clusters. On the other hand, the relative reconstruction error decreases monotonically with the number of latent features. This decrease is more prominent up to the estimated number of topics followed by a reduced change in the error. As observed in Figure \ref{fig:nmfk_sil_plots}, with the further increase in the number of latent features past the estimated $k$, there is a sudden decline in the Silhouette score due to the over-fitting phenomenon as the model tends to fit noise.

\subsection{Hierarchical Non-Negative Matrix Factorization}
\label{sec:hierarhicalnmf}

The NMF and Hierarchical NMF \cite{kuang2013fast, grotheer2020covid} strategies have been used successfully for document clustering \cite{xu2003document,carel2021simultaneous}, and topic modeling \cite{greene2014many,shi2018short, Vangara2020}. Here we use \textit{NMFk} to compute clusters of malware specimens by applying it in a hierarchical manner, where successive node expansions focus on the subset of $\mat{X}$ obtained from the parent cluster. Here the clusters are determined using the columns of $\mat{W}$ via \textit{W-clustering} that cluster the specimens' coordinates in the reduced space (i.e. the rows of the matrix $\mat{W}$) \cite{vangara2021finding}, a topic we address in Section \ref{sec:hNMFkclassifier}. When going deeper in the graph towards the leaves, we investigate different characteristics of the specimens in the same group, and achieve better separability of the malware specimens.

Let us consider a simple example from hierarchical document clustering. We assume three well-curated clusters in a text corpus of news articles about sports, technology, and the economy. If we cluster these documents with \textit{NMFk} and \textit{W-clustering} we can obtain three \textit{"super"} clusters for sports, technology, and the economy. We can further divide the cluster containing sport articles into sub-topics, such as soccer, football, tennis, skiing etc. by applying additional iterations of \textit{NMFk}. In our analysis, we choose to select back the specimens corresponding to each one of the super clusters and apply \textit{NMFk} again. This is the idea behind the hierarchical approach, and consequently a hierarchical approach is used in the \textit{HNMFk Classifier} to further separate more heterogeneous clusters based on the known (or labeled) malware instances.

In our semi-supervised setting, the experimental setup contains data with labeled (known) and unlabeled (unknown) malware specimens. This allows us to choose a scoring function, not based on information gain (such as normalized discounted cumulative gain from information retrieval \cite{10.1145/582415.582418}) \cite{kuang2013fast} or a fixed threshold using the number of specimens in the cluster \cite{grotheer2020covid} to determine which node to take further. Instead, we use a \emph{cluster uniformity score} that measures the stability of the cluster, based on the known specimens in the cluster, as the node expansion criteria for a cluster. We will further explain how we calculate the cluster uniformity score in Section \ref{sec:hNMFkclassifier}. In general, the application of \textit{NMFk} to semi-supervised data will place into each of the final clusters both labeled and unlabeled malware specimens. This allows us to continue to build the hierarchical graph until the further expansion of a particular node is stopped if no unlabeled or labeled samples are present in this node, or if the cluster uniformity score calculated based on the known samples passes the provided threshold.

\begin{figure}[htb]
    \centering
    \includegraphics[width=1\textwidth]{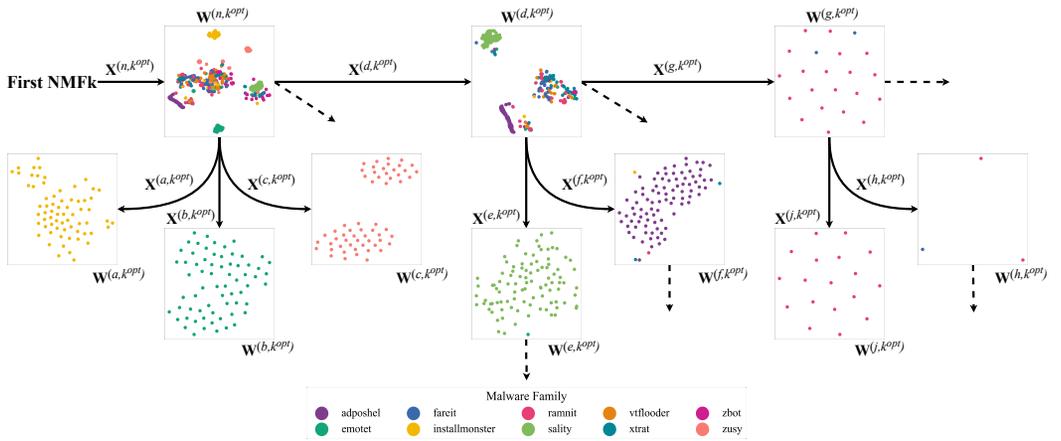}
    \caption{The path of the hierarchical graph formed by the NMFk is shown using 1,000 malware specimens containing a total of 10 malware families. After each factorization, the clustering is visualized by reducing the dimensions of $\mat{W}$ using t-SNE. Dashed arrows are used to indicate the existence of an another sub-tree from the node. Since we are obtaining $k_{opt_i}$ subsets of specimens from current $\mat{X}$ at each stage, $n > a \geq b \geq c \geq d > e \geq f \geq g > j \geq h$.}
    \label{fig:graph_tsne}
\vspace{-1.3em}
\end{figure}

We provide an example visualization of the latent factors obtained from \textit{NMFk} with a hierarchical setting in Figure \ref{fig:graph_tsne}. Here, we apply dimensionality reduction using \textit{t-SNE} \cite{vanDerMaaten2008} to each latent factor $\mat{W}$ to plot the clusters. Each point in the embedding of $\mat{W}$ is colored based on the family to which the specimen belongs. Here the clusters are expanded until all the samples in the cluster belong to a single class. The \textit{t-SNE} visualization show how the hierarchical clusters of malware families are formed, and how the clusters become more homogeneous as we perform additional applications of \textit{NMFk}.

\subsection{HNMFk Classifier}
\label{sec:hNMFkclassifier}

In this section, we describe how our \textit{NMFk} based hierarchical bulk classifier works. Our model recursively analyzes the known and unknown specimens, factorizing only the subset of data from the previous cluster at each iteration. 

The hyper-parameters of our model are the hyper-parameters needed for \textit{NMFk}, and the cluster uniformity threshold $t$. The user specifies the maximum number of iterations for NMF, number of perturbations, the error rate, and the range to search for the $k$ heuristic. When performing classification, \textit{HNMFk Classifier} is provided with the data matrix $\mat{X} \in \mathbb{R}_{+}^{n \times m}$, where $n$ is the number of malware samples and $m$ is the number of features, which includes both the known and unknown specimens that we want to perform inference on. We also provide a vector $\vect{y}$ containing the labels for each specimen. The $i$th sample, where $1 \geq i \geq n$, has the family label $\vect{y}_i \in \{-1, 1, 2, \dots, C \}$ for a dataset with $C$ classes. Notice that the unknown specimens are labeled with $-1$.

Our algorithm proceeds with the first factorization, given $\mat{X}$, $\vect{y}$, the specified \textit{NMFk} hyper-parameters, and cluster uniformity threshold $t$ as input. After \textit{NMFk} identifies the number of clusters $k^{opt}$, we obtain the latent factors $\mat{W} \in \mathbb{R}_{+}^{n \times k^{opt}}$ and $\mat{H} \in \mathbb{R}_{+}^{k^{opt} \times m}$. \textit{HNMFk Classifier} uses $\mat{W}$ latent factor to perform clustering, which we call \textit{W-clustering}. Here each $n$ sample is assigned to one of $k^{opt}$ clusters by taking the maximum value along the second axis:
\begin{equation}
    \label{eq:W_clustering}
    \text{cluster}(i) = \underset{0\leq j \leq k^{opt}}{\operatorname{arg\,max}} \, (\mat{W}_{ij})
\end{equation}
where $\text{cluster}(i)$ returns the cluster assignment of a given sample $i$. If a cluster $c$, where $c \in \{1, 2, \dots, k^{opt}\}$, does not contain any known samples, all the unknown specimens in the cluster $c$ are predicted abstaining. If a cluster $c$ only has known specimens, we do not proceed with the samples in that cluster further, as there are no more unknown specimens to label. On the other hand, if a cluster $c$ has a mix of known and unknown samples, we calculate the uniformity of the cluster based on the known specimens. Our cluster uniformity score is defined by the fraction of the most dominant class present in the cluster $c$:
\begin{equation}
    \label{eq:cluster_quality}
    U^c = \dfrac{|\operatorname{max} \, (c^{known})|}{|c^{known}|}
\end{equation}
where $U^c$ is the cluster uniformity score for the cluster $c$, $|c^{known}|$ is the number of known samples in the cluster, and the numerator is the number of samples that belongs to the most dominant known class in $c$. $U^c$ specifies how uniform the given cluster $c$ is based on the labeled data.

\begin{figure}[htb]
    \centering
    \includegraphics[width=1\textwidth]{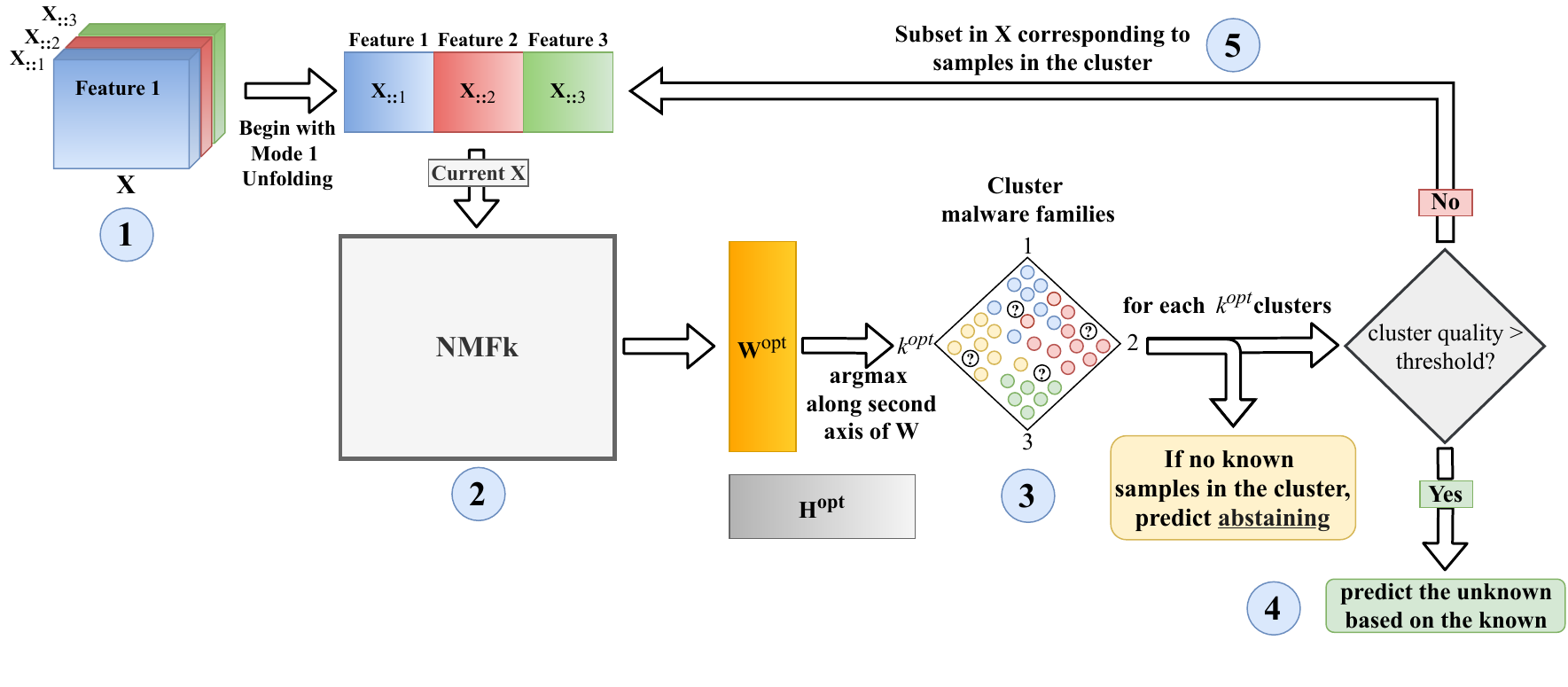}
    \caption{Overview of the HNMFk Classifier framework. NMFk is wrapped around an hierarchical (or recursive) semi-supervised architecture. Begin with the initial data $\mat{X}$ (1). Use NMFk to estimate the number of clusters and obtain the latent factor $\mat{W}$ (2). Extract the clusters via argmax along the second axis of $\mat{W}$ (3). For each cluster, perform abstaining prediction if no known samples are present in the cluster, or predict the unknown specimens in a semi-supervised manner if the cluster uniformity score is satisfied (4). Form the new matrices $\mat{X}$ with the specimens from the clusters that does not meet the cluster uniformity threshold (5). For each new $\mat{X}$, apply NMFk again (2).}
    \label{fig:hnmfk_overview}
\end{figure}

If the cluster uniformity score $U^c$ is more than the threshold $t$, then we proceed to assign unknown specimens family labels in a semi-supervised fashion. That is, all the unknown samples are predicted to be the most dominant class in the cluster based on the known specimens ($\operatorname{max} \, (c^{known})$). If, however, the cluster uniformity score is less than the threshold $t$ for a given cluster $c$, we form a new $\mat{X}^{'}\in \mathbb{R}_{+}^{|c| \times m}$ that only contains the malware specimens present in that cluster (both known and unknown). Finally, $\mat{X}^{'}$ is factorized again with \textit{NMFk}. In the proceeding \textit{NMFk}, $k$ search range selected to be $[1, k^{opt}]$ with the step-size of 1. The above procedure is repeated until all the unknown samples are classified. In this setting, our leaf nodes in the hierarchical graph are the positions where at least one of the following exit conditions are met: no known specimens in the cluster (abstaining prediction), no unknown specimens are in the cluster (nothing to classify), or $U^{c}\geq t$ is true and we classify all samples in the cluster in a semi-supervised manner. The aforementioned procedure is summarized in Algorithm \ref{algorithm:hnmfk} and Figure \ref{fig:hnmfk_overview}.

\begin{algorithm}[htb]
    \caption{HNMFk Classifier($\mat{X}$, $\vect{y}$, $k^{min}$, $k^{max}$, $r$, $t$) - Semi-supervised Hierarchical Classifier \label{algorithm:hnmfk}} 
	\begin{algorithmic}[1]
    	\State known\_samples= argwhere($\vect{y}$ != -1)
        \State unknown\_samples = argwhere($\vect{y}$ == -1)

		\State $\mat{W}$, $\mat{H}$, $k^{opt}$ = NMFk($\mat{X}$, $k^{min}$, $k^{max}$, $r$)
		\State clusters = argmax($\mat{W}$, axis=1)
		
		\For {cluster in clusters} \Comment{iterate over $k^{opt}$ clusters}
		
		    \State known\_samples\_c = intersect(known\_samples, cluster)
            \State unknown\_samples\_c = intersect(unknown\_samples, cluster)
		    
		    \If {len(known\_samples\_c) == 0} \Comment{no unknown samples to make prediction}
		       \State \textbf{continue}
		    \EndIf
		    
		    \If {len(unknown\_samples\_c) == 0} \Comment{\textit{abstaining} prediction}
		       \State \textbf{continue}
		    \EndIf
		    
		    \State class\_counts = count(known\_samples\_c)
		    \State cluster\_uniformity = max(class\_counts) / sum(class\_counts)
		    
		    \If {cluster\_uniformity < t}
		        \State $\mat{X}$\_new = $\mat{X}$[cluster] \Comment{subset in $\mat{X}$, samples in the cluster}
		        \State $\vect{y}$\_new = $\vect{y}$[cluster] \Comment{labels for the samples in the cluster}
		        \State $k^{max}$ = min($k^{opt}$+1, min($\mat{X}$.shape))
		        \State y[cluster] = HNMFk\_Classifier($\mat{X}$\_new, $\vect{y}$\_new, $k^{min}$, $k^{max}$, r, t)
		        
		    \Else
		        \State classify\_label = max(class\_counts) \Comment{dominant known class in the cluster}
		        \State y[unknown\_samples\_c] = classify\_label

            \EndIf
            
		\EndFor
	
	\State \textbf{return} $\vect{y}$
	
	\end{algorithmic} 
\end{algorithm}

In summary, looking at Figure \ref{fig:hnmfk_overview} we can conclude that \textit{HNMFk Classifier} is a wrapper to the \textit{NMFk} algorithm, which exploits \textit{NMFk}'s ability to estimate the number of latent components, and performs factorization recursively to create a hierarchical graph where the semi-supervised classification is done at each leaf node. When our model finishes classification, any unknown samples that are left with the label $-1$ are said to be abstaining predictions, i.e. the model does not know their classes or rejects to make a prediction.

\begin{figure}[htb]
    \centering
    \includegraphics[width=1\textwidth]{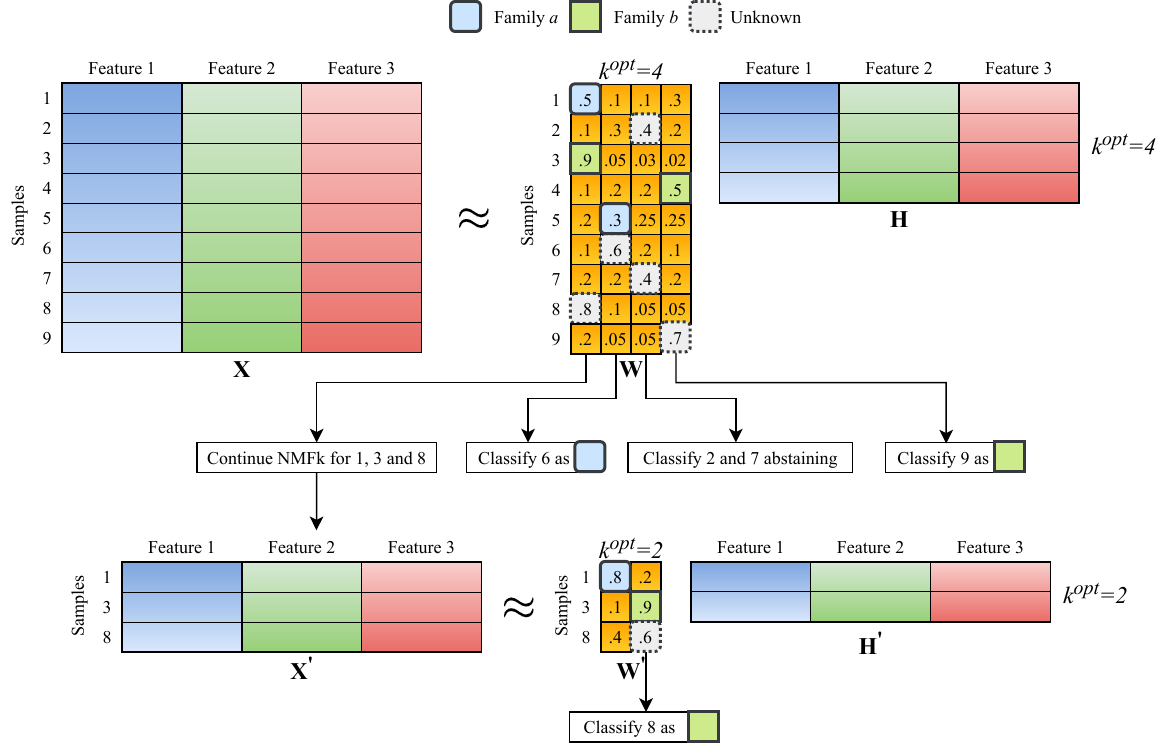}
    \caption{A toy demonstration of how HNMFk Classifier operates in a hierarchical fashion, and how the semi-supervised classification of the unknown malware specimens is performed via the clustering on the latent $W$ matrices using the known samples.}
    \label{fig:hnmfk}
\vspace{-1.2em}
\end{figure}

We also provide a toy example illustrating how \textit{HNMFk Classifier} works in Figure \ref{fig:hnmfk}. In this figure, we have a matrix $\mat{X} \in \rm I\!R^{9\, x\, 3}$ (9 malware samples with 3 features). After factorizing $\mat{X}$ with \textit{NMFk}, we get the latent factors $\mat{W} \in \mathbb{R}_{+}^{9 \times k^{opt}}$ and $\mat{H} \in \mathbb{R}_{+}^{k^{opt} \times 3}$, with the estimated number of clusters $k^{opt}=4$. Samples 5 and 6 are assigned to cluster 2. Sample 6, an unknown sample, is classified as family $a$. Cluster 3 contains only 2 unknown samples. Therefore, we classify samples 2 and 7 as abstaining. Cluster 1 contains the sample 1 (family $a$), 3 (family $b$), and 8 (unknown). Because this cluster has samples belonging to two different families (assuming that our cluster uniformity threshold is $t=1$, i.e. threshold is met only when all the known samples in the cluster belongs to a single class), we create a new subset with these samples, such that $\mat{X}^{'} \in \mathbb{R}_{+}^{3 \times 3}$. We apply \textit{NMFk} again on $\mat{X}^{'}$, which estimates $k^{opt}=2$, and sample 8 is classified as family $b$. When all samples are predicted, the computation is complete.

\section{Dataset}
\label{sec:dataset}

Collection of malware data has challenges such as copyright issues, labeling difficulty, and security precautions. Therefore, compared to other ML fields with abundant data (such as text and images), the malware identification community has lacked a benchmark dataset sufficient to enable reproducibility and comparison of new methods. To address this issue, Anderson et al. released the EMBER-2018 dataset \cite{Anderson2018}, which we use in our experiments. Since its release, EMBER-2018 has become a popular benchmark dataset for ML-based malware analysis methods. 

EMBER-2018 is a collection of PE header and meta-data information extracted from 1.1 million benign and malicious Microsoft Windows Portable executable binaries, out of which 800,000 have labels. The family labels in the dataset are obtained using AVClass. Therefore, classes are \textit{weakly labeled} as AVClass contains inaccuracies in family labeling \cite{8871171}. AVClass does not filter out all generic family names, it can be inconsistent in its use of aliases for malware families, and errors in any antivirus signatures can effect AVClass' accuracy. Despite the imperfect labeling, AVClass is currently the best available option for obtaining a large quantity of malware family labels.


Throughout our analysis, we only use the malware instances for which AVClass could determine a family label. The final dataset includes a default train and test split, where the training set consists of over 289,000 specimens from 2,730 malware families, and the test portion of the dataset includes around 99,000 samples from 916 malware families. The details of the dataset sample and family statistics are shown in Table \ref{table:dataset_stats}. In this paper, we refer to the training set as \textit{known} data, and testing set as \textit{unknown} data in the context of semi-supervised learning and bulk classification (i.e. we use the \textit{known} data as a reference to label the \textit{unknown} data).

\begin{table*}[htb]
\caption{EMBER-2018 dataset default train and test set split and malware family and sample counts are displayed. Novel families for the known (or train) set are the families that only exist in the training set. The novel families for unknown (or test) set are the families that only exist in the test set (i.e. we do not see these families during inference, or we do not have known specimens for reference). Min Family and Max Family columns show the minimum and maximum number of samples exist for a family in the dataset. For instance, there are malware families with single sample in both known and unknown sets. Samples/Family column shows the average number of samples per family. We used all the malware instances with a family label from EMBER-2018, which contains the rare and novel families, making the classification task complex.}
 \label{table:dataset_stats}
\resizebox{\columnwidth}{!}{
\begin{tabular}{l|c|c|c|c|c|c|c}
\hline
\textbf{Set}                 & \textbf{Families}     & \textbf{Samples} & \textbf{Novel Families} &  \textbf{Novel Samples} &  \textbf{Min Family} &  \textbf{Max Family} &  \textbf{Samples/Family}\\ \hline

\text{Known (Train)}    & 2,730 & 289,026 & 1,982 & 11,157 & 1 & 16,689 & 105.87\\
\text{Unknown (Test)}    & 916  & 99,216 & 168 & 363 & 1 & 19,260 & 315.53\\

\hline
\end{tabular}
}
\end{table*}

\begin{figure}[htb]
\centering
\includegraphics[width=1\textwidth]{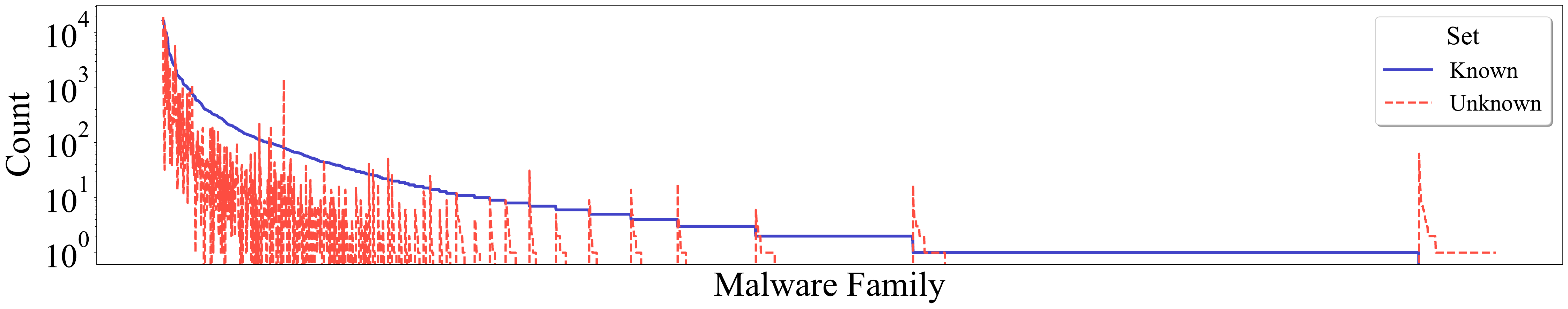}
\caption{Distribution of the malware families in EMBER-2018 dataset. Count of family classes are shown in log scale for both the known and unknown set. Both the known and unknown sets has an extremely imbalanced classes of malware families, and the unknown set of contains set of novel malware families. \label{fig:family_dist}}
\vspace{-1.2em}
\end{figure}

One advantage of using the EMBER-2018 dataset is that the distribution of the family classes resembles real-world cases. The \textit{known} portion of the dataset contains malware families that do not exist in the \textit{unknown} portion of the data. Similarly, the \textit{unknown} set contains novel malware families, or the malware families that do not exist in the \textit{known} set. This is also shown in Table \ref{table:dataset_stats}. 1,982 of the malware families, making over 11 thousand samples, are not seen again in the \textit{unknown} set. There are 168 novel families, forming 363 samples, that we do not have any reference of in the \textit{known} set. At the same time, malware family classes in EMBER-2018 are extremely imbalanced. Figure \ref{fig:family_dist} shows the distribution of the malware families for both the \textit{known} and \textit{unknown} set. For instance, there are malware families that consist of single samples, including the specimens from the novel families (which can also be seen at the right side of Figure \ref{fig:family_dist} with red-dashed line). In fact, the majority of the malware families in the dataset consist of less than 10 samples. We next proceed to the pre-processing of the features to remove the outliers. 

\subsection{Pre-processing}
\label{sec:pre_processing}

During our experiments, we represent each file in the dataset as a collection of features from both general file meta-data as well as PE header information. Each of the features are concatenated horizontally to form the final features matrix. This is equivalent to forming an 11 dimensional tensor, with the dimensions \textit{Samples $\times$ Feature 1 $\times$ Feature 2 $\times$ ... $\times$ Feature 10}, and taking the mode-1 unfolding of the tensor. Specifically, we use the following features:
\begin{enumerate}
    \item \textit{byte histogram}: a vector of size 256 where each entry represents the number of times a certain byte occurs in the file.
    \item \textit{byte entropy}: normalized joint distribution of entropy and byte values.
    \item \textit{print table distribution}: distribution of characters obtained from printable strings with minimum of 5 consecutive printable characters in the binary.
    \item \textit{strings entropy}: measure of randomness of printable strings present in the malware.
    \item \textit{number of strings}: number of printable strings.
    \item \textit{file size}: size of the binary in bytes.
    \item \textit{number of exports}: number of functions exported by the malware.
    \item \textit{number of imports}: number of functions imported by the malware.
    \item \textit{code size}: size of \textit{.text} or code section of the PE header in bytes.
    \item \textit{number of sections}: number of sections present in PE header.
\end{enumerate}

Our dataset consists of heterogeneous features containing outlier values. Since NMF is susceptible to outliers (extremely large or small values in the columns of initial data), see for example \cite{zhang2011robust}, we normalize the features used in our analysis. This normalization prevents the larger values in  the columns of the initial data to bias/skew the NMF optimization procedure by favorizing some of the columns in X, see details in Ref.\cite{islam2022uncovering}. Note that the case of outliers affecting  NMF optimization is distinct from the characteristics makeup of a novel malware family: After the normalization, the novelty of the malware belonging to unknown (or never seen before) family reflects on the shape of its latent signature (the columns of matrix W). A possible classification of malware families based on their latent signatures will be discussed elsewhere.

\begin{figure}[htb]
\centering
\includegraphics[width=1\textwidth]{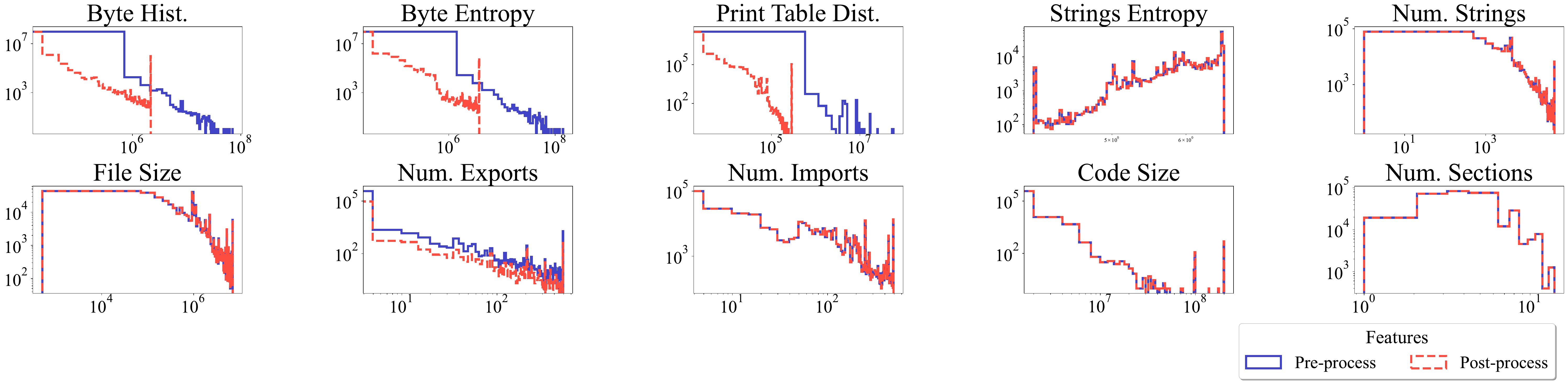}
\caption{Static malware analysis based features from PE header files and malware meta-data used in the analysis shown before and after the mapping of the outliers, defined by $Z=3$ statistical score, for both training (known specimens) and test sets (unknown specimens). \label{fig:normalization}}
\vspace{-1.2em}
\end{figure}

In our normalization, Z-scores are used to remap the outliers that are more than or less than 3 standard deviations away from the mean. These outliers are mapped to the point that is exactly 3 standard deviations away from the mean. In Figure \ref{fig:normalization}, we show the histogram of feature values for pre- and post-processing. The normalization was most prominent among the features \textit{byte histogram}, \textit{byte entropy}, and \textit{print table distribution}. Finally, we scale the values to be between 0 and 1.

\subsection{Preparation of the Experiments}
\label{sec:experiment_preperation}
We conduct our experiments using two different dataset setups. With the first setup, we use a subset of data utilizing only the top populous malware families to perform performance analysis of our method under different conditions in Section \ref{sec:performance_analysis}. This setup is also used in our ablation studies in Section \ref{sec:ablation_studies}. We use a smaller subset of the data to reduce the computation time of our experiments and to enable testing of our method under number of different settings. Although this setup allows us to gain insights into how our method works, it does not yield results that can generalize to real-world. Therefore, in Section \ref{sec:large_scale_classification} we test our method under realistic conditions and compare to other baseline models.

In our small dataset setup, we chose the 10 most populous malware families in the entire dataset (adposhel, emotet, fareit, installmonster, ramnit, sality, vtflooder, xtrat, zbot, zusy). We then randomly sample the dataset to extract 1,000 specimens for each family without replacement, forming a small subset of the dataset with 10,000 samples. We form 10 of these random subsets, and apply our experiments on each of the 10 subsets, to see if our results are statistically significant. In our large scale analysis, we used all of the malware families present in the EMBER-2018 dataset, and use the default split present in EMBER-2018 to separate the known and unknown specimens.

\subsection{System Configuration}
We ran the experiments on a High Performance Computing (HPC) cluster named Dracarys, 
located at the Los Alamos National Laboratory (LANL). 
Dracarys 
uses the Intel(R) Xeon(R) Platinum 8280M processor, which is a cascade lake architecture operating at a clock speed of 2.70GHz. There are 28 physical CPU cores which are multi-threaded to 56 threads providing 112 virtual processors, and total physical RAM of 2.71 TeraBytes (TBs). The system also comprises 3 NVIDIA Quadro RTX8000 GPUs with VRAM memory of 48 GigaBytes (GBs) each.

\section{Experiments}
\label{sec:results}

We perform experiments targeting the following tasks: analyzing the performance of our model with different hyper-parameters and, as the amount of known malware decreases, and testing our method under realistic conditions. We compare our results to those obtained by the baseline models, taking advantage of the abstaining prediction ability to detect novel malware, and using ablation studies to justify the need for the parts of our model.

\subsection{Methodology Performance Analysis}
\label{sec:performance_analysis}

In this section we look at the performance of our method for different cluster uniformity thresholds, unknown malware fractions, and \textit{NMFk} hyper-parameter selections. Similar to prior work, we use a small subset of the dataset (an unrealistic data setup), as described in Section \ref{sec:experiment_preperation}, during our analysis in this section. Each experiment is run 10 times on different random subsets of the dataset, to verify if the results are statistically significant using hypothesis testing. To this end, we report our results with a 95\% confidence interval (CI) for each experiment.

\subsubsection{Cluster Uniformity Threshold}
\label{sec:cluster_quality}

\begin{figure}[htb]
\centering
\includegraphics[width=1\textwidth]{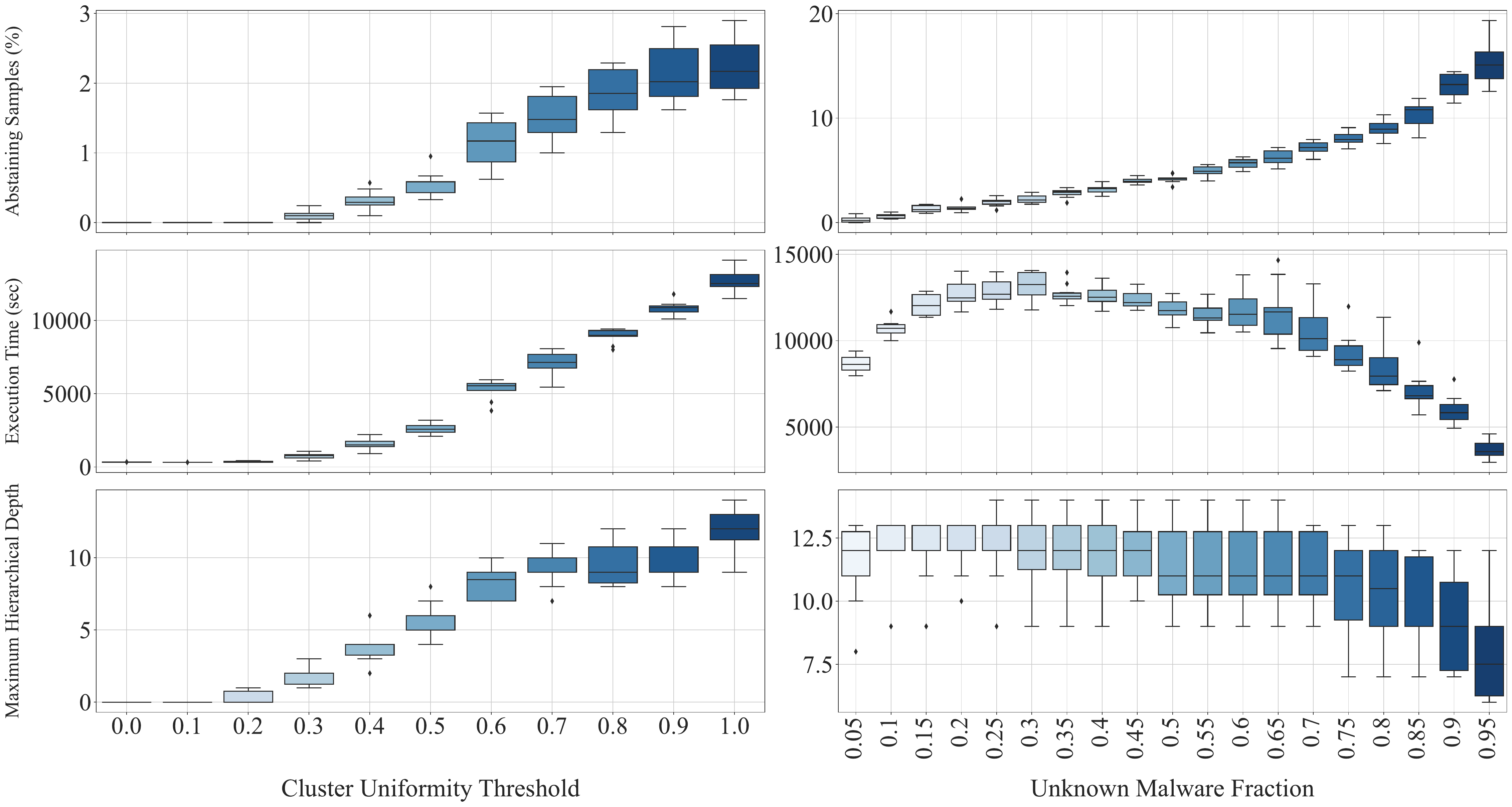} %
\caption{HNMFk Classifier's performance for abstaining prediction, execution time, and the maximum depth is shown as the cluster uniformity and the unknown malware fraction changes.  \label{fig:methodology_analysis}}
\end{figure}

\begin{figure}[htb]
\centering
\includegraphics[width=1\textwidth]{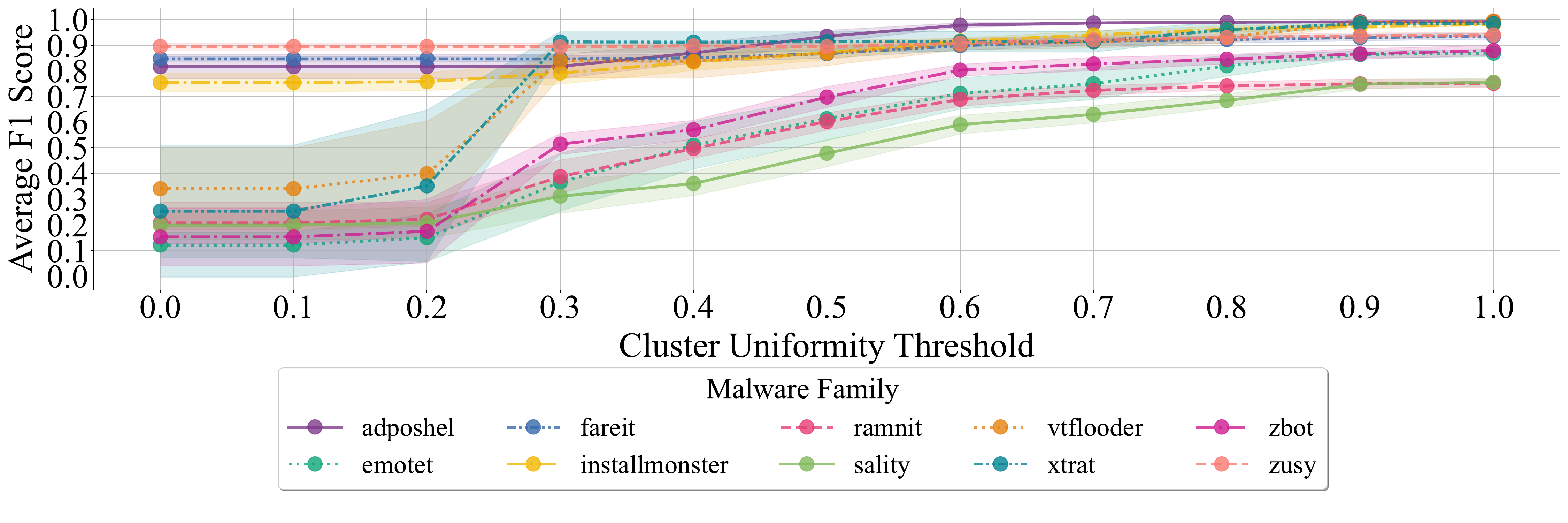}
\caption{The performance of HNMFk Classifier is measured with the F1 score as 
the cluster uniformity threshold is changed. Each experiment is performed on 10 different random subset of the EMBER-2018 dataset, average is plotted with the 95\% confidence interval. \label{fig:validation_set_results}}
\end{figure}

We use a threshold value $t$, which measures how many labeled (known) specimens are needed to claim that all unknown specimens in this cluster are uniform, that is, from the same labeled malware family. This threshold allows us to determine whether to proceed further with clustering of the current data in the node with additional applications of \textit{NMFk}. The left side of Figure \ref{fig:methodology_analysis} shows the percent of abstaining predictions, execution time, and the maximum graph depth (maximum number of edges between the root and a leaf node) as the cluster uniformity threshold $t$ is changed. As $t$ increases, the percent of abstaining predictions rises, since the solution needs increasingly cleaner clusters. This reduces the number of specimens that we can classify with high certainty, and results in a higher number of abstaining predictions. The maximum graph depth also increases, alongside the higher execution time, since achieving cleaner clusters requires an increased number of separations. We show how the F1 score changes for each malware family in Figure \ref{fig:validation_set_results}. As the cluster threshold increases, the performance of the model improves for each malware family, and the results become more certain, as indicated by the narrowing confidence interval. Although the computation time increases, a higher threshold yields better inference results. Therefore, during the experiments in Section \ref{sec:large_scale_classification} we set the threshold to be $t=1$.

\subsubsection{Unknown Malware Fraction}
\label{sec:unknown_malware_fraction}
\begin{figure}[htb]
\centering
\includegraphics[width=1\textwidth]{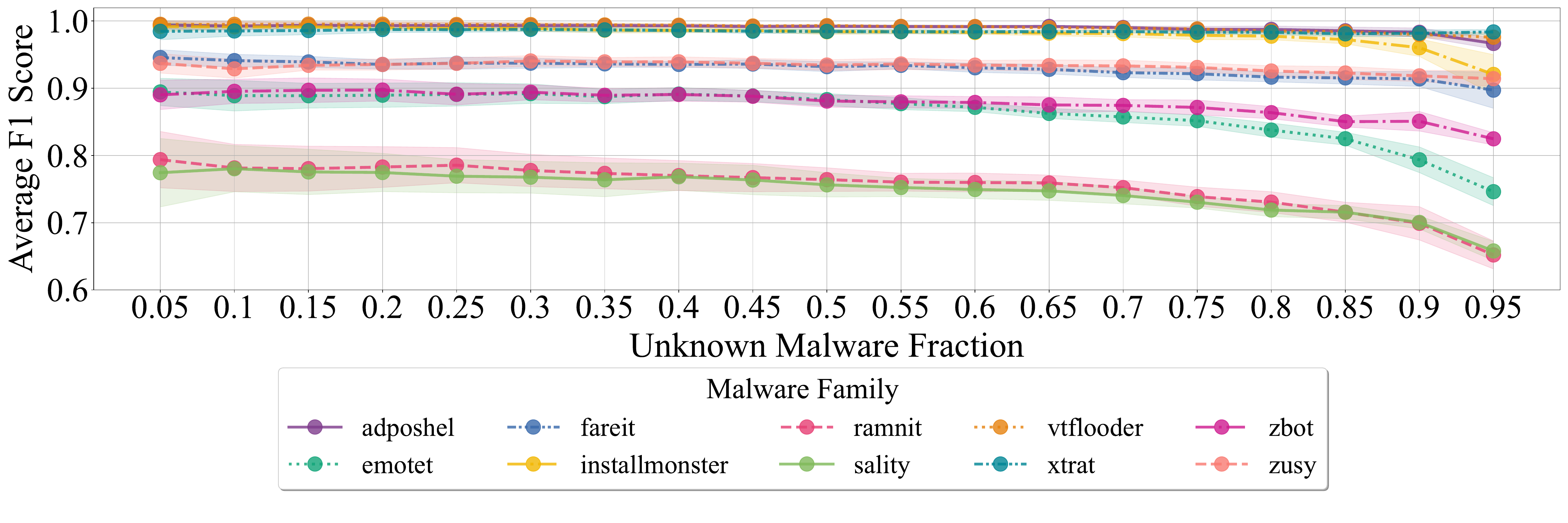}
\caption{The performance of the HNMFk Classsifer, measured with F1 score, remains relatively stable for each malware family as the unknown malware fraction increases (or the number of known samples decreases). Each experiment is run on 10 random subset of the dataset.\label{fig:fraction_malware_fa}}
\vspace{-1.2em}
\end{figure}

The process of labeling malware is expensive \cite{Raff2020ASO}; therefore, semi-supervised learning can help with obtaining good performance results when using a low quantity of labeled data. We investigate this by looking at how our model performs as the unknown malware fraction increases. Figure \ref{fig:fraction_malware_fa} displays the average F1 score for each malware family as the unknown malware fraction rises. Since our model can perform abstaining predictions, as the unknown malware fraction increases, the performance of the model remains relatively stable. A lesser number of known malware samples means that our model to have a lesser number of references that can be used to classify the unknown samples. This results in higher number of abstaining predictions which in return helps with maintaining the performance (this can be seen at the right top of Figure \ref{fig:methodology_analysis}). In Figure \ref{fig:fraction_malware_fa}, we can also see that two malware families, Sality and Ramnit, yield lower F1 scores in comparison to the other families. Possible reasons for diminished performance on Sality and Ramnit include the fact that they are both \textit{``file infectors''} (a category of malware which copies its code into other executables). It may be more difficult to classify this type of malware using the selected features, since some of the original PE metadata/file contents may not be changed when a file is infected.

\begin{figure}[htb]
\centering
\includegraphics[width=1\textwidth]{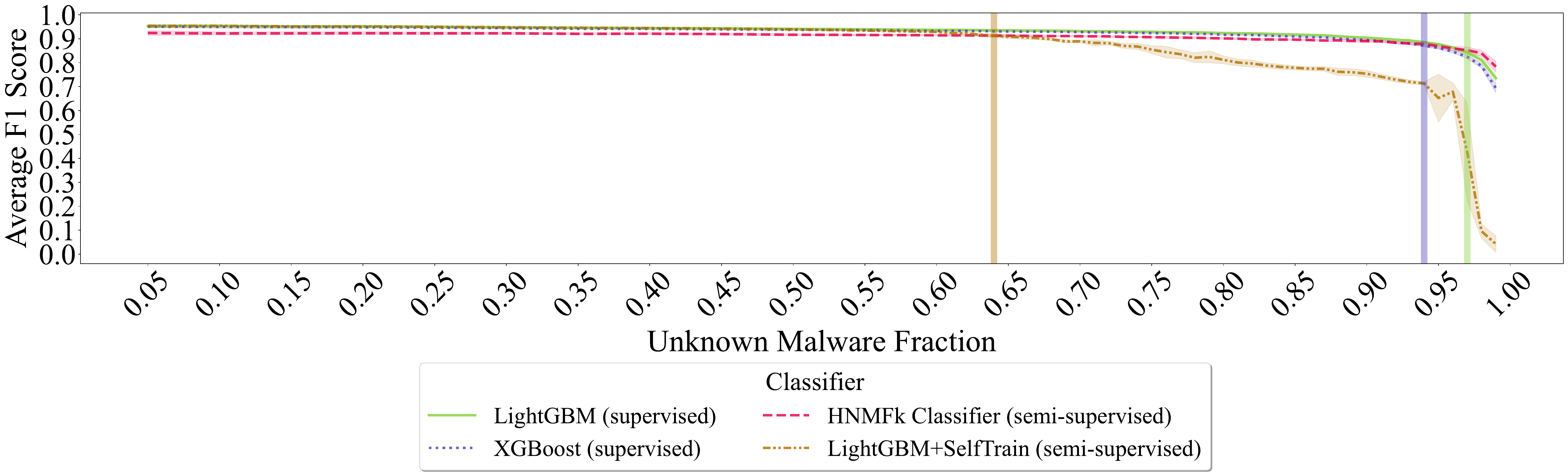}
\caption{Average F1 score when classifying 10 malware families is compared to other baseline models as the fraction of unknown malware increases. Each experiment is run on 10 random subset of the dataset.\label{fig:fraction_malware_fa_classifiers}}
\end{figure}

The average F1 scores obtained by the \textit{HNMFk Classifier} with the changing unknown specimen fraction are also compared to the vanilla baseline models in Figure \ref{fig:fraction_malware_fa_classifiers}. Here, the unknown malware fraction point where the \textit{HNMFk Classifier} begins to outperform a baseline model is shown with a vertical line. We use the supervised baseline models \textit{XGBoost} and \textit{LightGBM}, and a semi-supervised model \textit{LightGBM+SelfTrain}. These traditional ML models do not have the ability to perform abstaining predictions. Therefore, they rely on an abundance of labeled data to perform well during testing. The \textit{HNMFk Classifier} surpasses the average F1 score of \textit{LightGBM+SelfTrain} at 0.64 unknown malware fraction. \textit{XGBoost} is outperformed at unknown malware fraction 0.94, and \textit{LightGBM} at 0.97.  We also note that these models continue to perform relatively well as the known malware fraction drops because we are using a small and balanced subset of the dataset which contains the most populous malware families, making the problem easier. We will be further analyzing the performance of the baseline models and our approach with a realistic dataset setup in Section \ref{sec:large_scale_classification}. The experiments under real-world like setup will reveal that the performance difference between the baseline models and our method is even greater.

\subsubsection{NMFk Hyper-parameter Analysis}
\label{sec:nmfk_hyperparameter_selection}

\begin{figure*}[htb]
\centering
    \begin{minipage}[b]{0.31\linewidth}
        \includegraphics[width=1\linewidth]{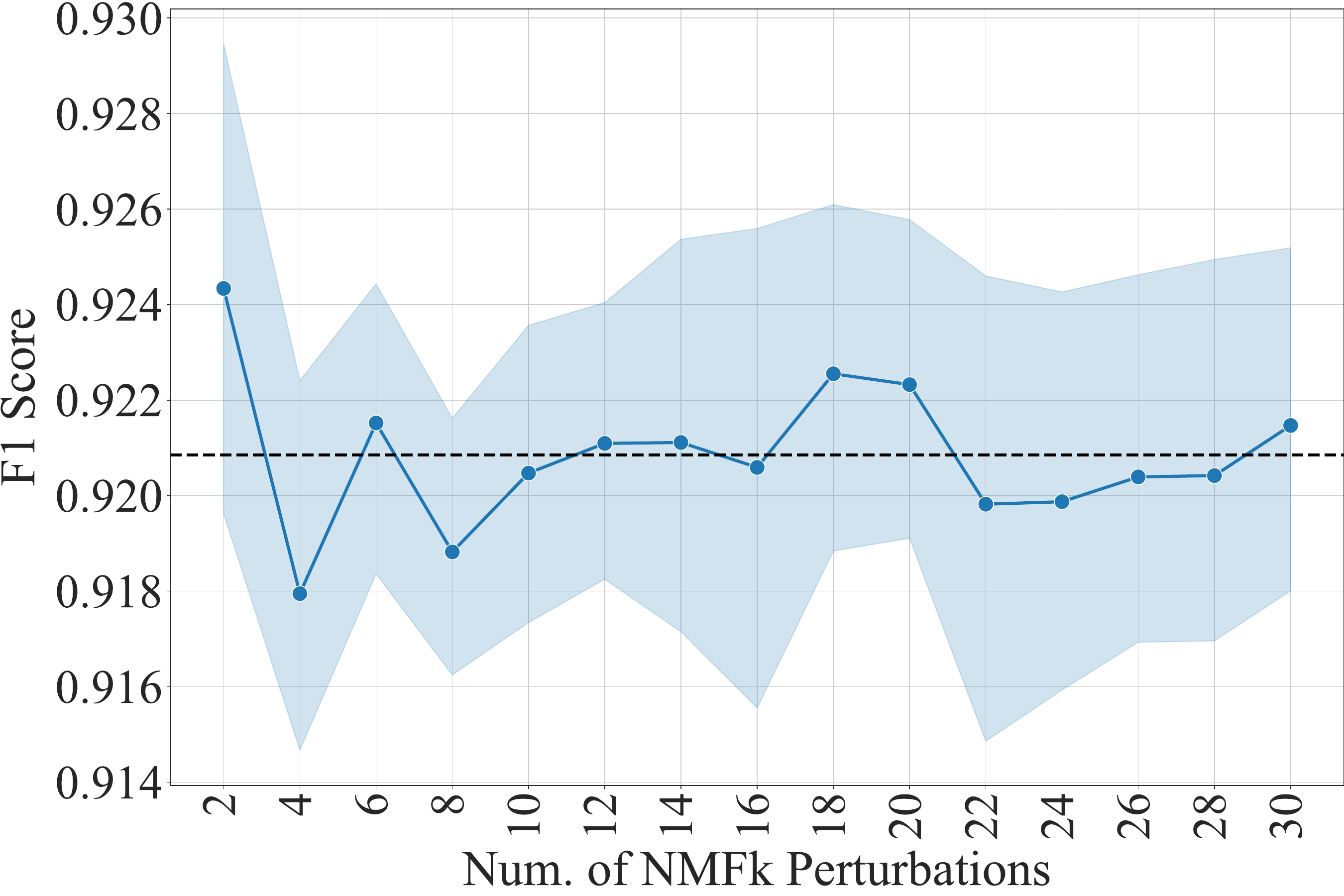}
        \caption{Change in performance is measured using F1 score as the number of NMFk perturbations increased. It can be seen the effect to the overall performance as this hyper-parameter is changed is low, with average average F1 score of .92 and confidence interval .001. The difference between the highest and lowest point is .032.}
        \label{fig:num_perturbation}
    \end{minipage}
    \quad
    \begin{minipage}[b]{0.31\linewidth}
        \includegraphics[width=1\linewidth]{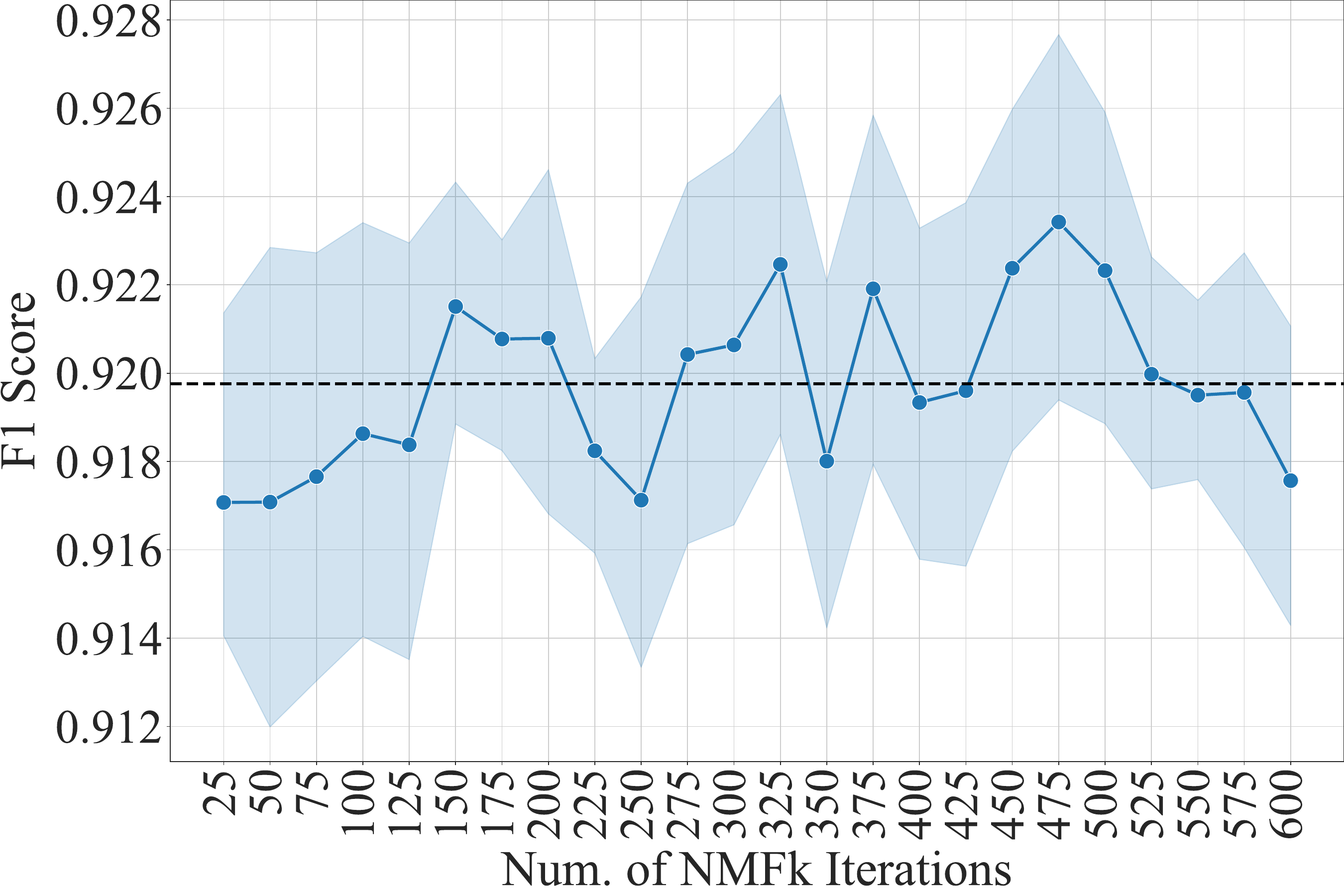}
        \caption{Change in performance is measured using F1 score as the number of NMFk iterations increased. It can be seen the effect to the overall performance as this hyper-parameter is changed is low, with average average F1 score of .91 and confidence interval .0008. The difference between the highest and lowest point is .031.}
        \label{fig:num_iterations}
    \end{minipage}
    \quad
    \begin{minipage}[b]{0.31\linewidth}
        \includegraphics[width=1\linewidth]{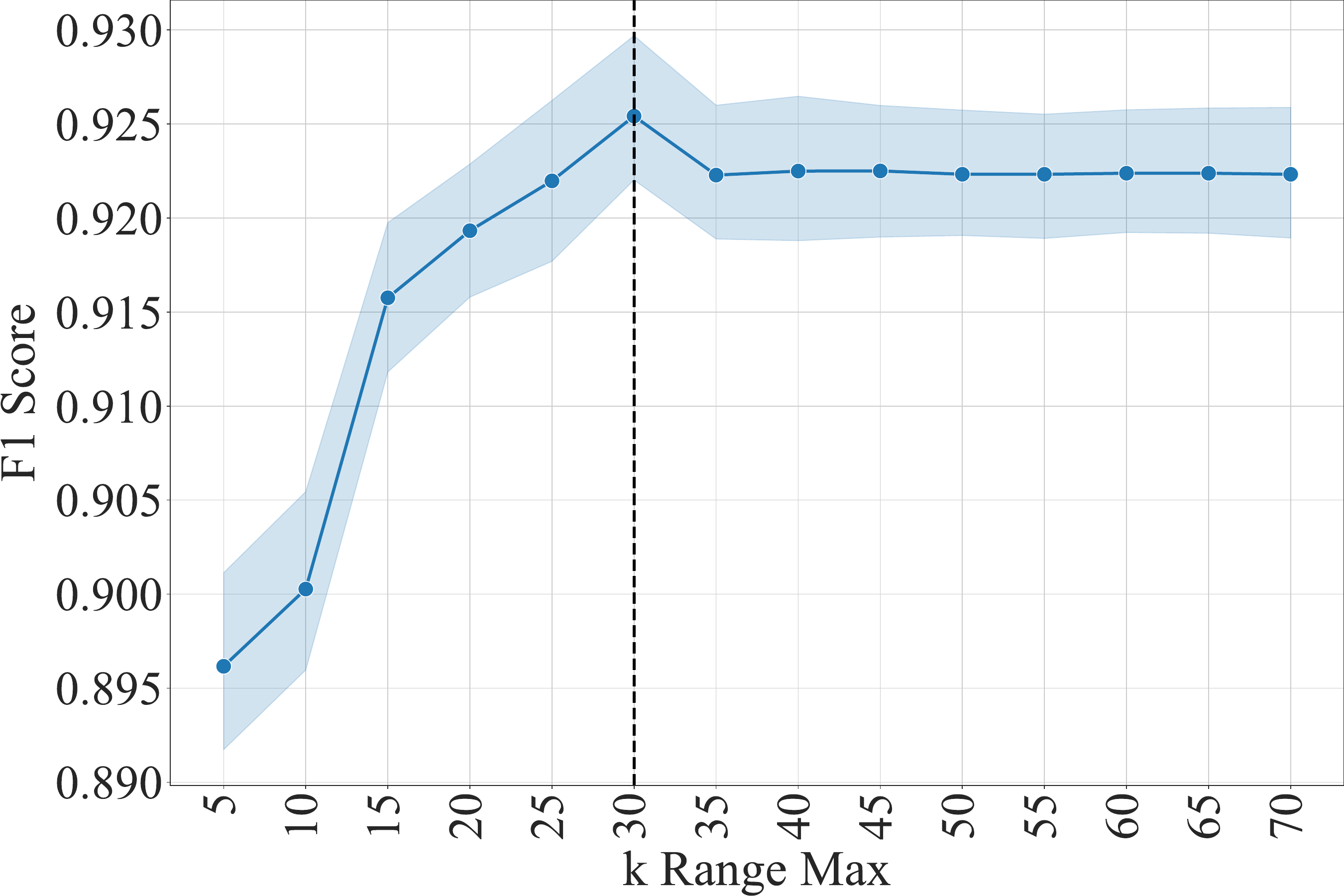}
        \caption{Change in performance is measured using F1 score as the maximum k for of NMFk k search is increased. Here, k range is 1 through the maximum k value, with the step size of 1. It can be seen as the maximum k increases, the performance of the model improves. After the estimated k, increasing the value further does not change the performance.}
        \label{fig:krange}
    \end{minipage}
    \vspace{-1em}
\end{figure*}

In addition to the cluster uniformity threshold hyper-parameter of the \textit{HNMFk Classifier}, we also provide our model with the hyper-parameters of \textit{NMFk}. In Figures \ref{fig:num_perturbation} and \ref{fig:num_iterations} we show that changes in the number of perturbations and NMF iterations do not have a large effect on the performance of our method. Figure \ref{fig:krange} displays the change in F1 score as the maximum $k$ is increased for the $k$ search of first \textit{NMFk}. In this experiment, we choose the $k$ step-size of 1, and begin searching at $k=1$. The performance of the model continues to increase as the predicted $k$ is approached. After the estimated $k^{opt}$ is reached, the F1 score does not change, since we will always choose the same $k^{opt}$ in the first \textit{NMFk}. These experiments indicate that we need to choose the initial $k$ search range to be large enough to obtain a good initial clustering.

\subsection{Malware Family Classification Under Realistic Conditions}
\label{sec:large_scale_classification}

Now that we have gained understanding into how our method performs with different hyper-parameters and settings, we will next use the more realistic data setup to show how our approach fares far better under real-world constraints. When ML-based malware defense and analysis solutions are used outside the research environment, they often encounter extreme class imbalance. At the same time, analysts do not have access to all possible malware samples, and threat actors continuously develop new pieces of malware. Therefore, ML-based systems are exposed to malware that has never been seen before. To this end, we analyze the performance of our method under a real-world like setting by exposing our model to prominent, rare, and novel malware families. In this section, we utilize all the malware families present in the EMBER-2018 dataset to conduct our experiment, as described in Section \ref{sec:experiment_preperation}. The performance of \textit{HNMFk Classifier} is compared to the supervised baseline models \textit{LightGBM}, \textit{XGBoost}, and \textit{MLP}. We also create strong semi-supervised versions of \textit{LightGBM} and \textit{XGBoost} by wrapping them with \textit{SelfTrain}. During the hyper-parameter tuning of \textit{LightGBM} and \textit{XGBoost}, we use the Python package \textit{Optuna} to get the hyper-parameter suggestions for each trial \cite{akiba2019optuna}, and for the construction of an optimal neural net-based classifier, \textit{MLP}, we employed a \textit{HyperBand Tuner} as an accelerated tuning algorithm \cite{JMLR:v18:16-558}.
The hyper-parameters of \textit{LightGBM} was tuned using a stratified 20\% subset of the training set over 65 trials and 3-fold stratified cross-validation. We used a stratified subset of the dataset because using the entire dataset for this model resulted in each trial taking approximately 2 days during tuning (it would have taken approximately 100 days to complete 50 trials for tuning). We used the objective \textit{multiclass} with a 500 maximum number of iterations, and \textit{gbdt} boosting type. The following hyper-parameters were tuned (ranges are shown in parenthesis): \textit{min\_data\_in\_leaf} (5-100 in log scale), \textit{max\_depth} (2-7), \textit{bagging\_freq} (0-5), \textit{bagging\_fraction} (.5-1.0), \textit{learning\_rate} (.001-.1 in log scale), and \textit{feature\_fraction} (.1-.7). For \textit{LightGBM}, we have also tried the recommended hyper-parameters from the EMBER-2018 dataset \cite{Anderson2018}, which did not yield better results when compared to our best trained model.

\textit{XGBoost} was tuned using the entire dataset over 25 trials with stratified 3 fold cross-validation. We used maximum boosting rounds of 500 with the multi-class softmax objective function. The following hyper-parameters were tuned: \textit{max\_depth} (2-10), \textit{eta} (.003-0.5 in log scale), \textit{subsample} (.2-.7), \textit{rounds} (10-300), \textit{colsample\_bytree} (.3-1.0), \textit{colsample\_bylevel} (.5-1.0), and \textit{lambda} (.1-2.0).

The \textit{HyperBand} framework has been widely used in the deep learning community for estimating the optimal parameters in a short amount of time. HyperBand is a variation of random search with explore-exploit theory to estimate best configurations within a given allocated time. The hyper-parameters utilized for model selection of the \textit{MLP} were the number of depths of the neural network (1-10), number of nodes on each layer (1024-16000), optimization algorithm (SGD, Adam, RmsProp), and the learning rate (1e-4, 1e-1). We employed early stopping criteria on validation loss to avoid over-fitting.

Since our method's performance does not change dramatically with the change in hyper-parameters as shown in Section \ref{sec:nmfk_hyperparameter_selection}, we choose the hyper-parameters without tuning with 20 perturbations, 500 number of iterations, and k-range to be 1 through 100 with the step-size of 1 for the first iteration. We did verify, by inspecting the plot of the initial \textit{NMFk} (similar to the Figure \ref{fig:nmfk_sil_plots}), that the estimated number of components was less than 100. If it had been close to 100, we would have re-started our experiment with a higher range. Finally, we chose the cluster uniformity threshold $t$ to be 1, i.e. each cluster should have a single known class to be able to perform semi-supervised classification.

Table \ref{table:baseline_comparisons_scores} compares our method to the baseline models. The \textit{HNMFk Classifier}, a semi-supervised solution, outperforms all of the state-of-the-art models, which we used as baselines, with an F1 score of 0.80. Our approach outperforms the supervised methods, with the potential benefit of better generalizability and the need for less labeled data, due to the semi-supervised setting. We also surpass the strong semi-supervised version of \textit{XGBoost} with \textit{SelfTrain}. Notice that these baseline models were used to report benchmarks by prior studies. However, our experiment reveals the performance of these models under realistic conditions. 

\begin{table*}[htb]
\caption{HNMFk Classifier is compared against the state-of-the-art supervised classifiers. HNMFk Classifier, a semi-supervised method, surpasses the previous state-of-the-art models, which are supervised, in malware family classification. \textbf{Weighted} F1, Precision, and Recall scores are provided for multi-class classification with imbalanced data. The F1 scores of HNMFk Classifier and HNMF2 Classifier does not include the abstaining predictions (score includes the specimens where the prediction was not rejected)}.
 \label{table:baseline_comparisons_scores}
\resizebox{\columnwidth}{!}{
\begin{tabular}{l|c|c|c|c|c}
\hline
\textbf{Model}                    & \textbf{F1}     & \textbf{Precision} & \textbf{Recall} &  \textbf{Tune Time}  &  \textbf{Train\&Predict Time} \\ \hline
\text{HNMFk Classifier (semi-supervised)}        & \textbf{0.80}      & \textbf{0.85}            & \textbf{0.77}       & 5.77 days & 7.91 days \\
\text{HNMF2 Classifier (semi-supervised, ablation study)}       & 0.77            & 0.82         & 0.74       & 5.77 days & 2.83 days \\
\text{XGBoost+SelfTrain (semi-supervised)}    & 0.76              & 0.78             & 0.73            & 2.06 days & 4.72 hours         \\
\text{XGBoost (supervised)}                     & 0.74           & 0.77              & 0.72            & 2.06 days & 2.93 hours\\   
\text{LightGBM (supervised, tuned on stratified subset)}        & 0.65           & 0.74              & 0.64            & 11.09 days & 3.02 hours \\ 
\text{MLP (supervised)}                         & 0.72               &  0.76              &  0.71            & 1.02 days & 30 minutes          \\
\text{LightGBM+SelfTrain (semi-supervised)}    & 0.64               & 0.69              & 0.61            & 11.09 days & 9.44 hours         \\
\hline
\end{tabular}
}
\vspace{-1.2em}
\end{table*}

Additionally, our method utilizes abstaining predictions (rejection to make a prediction), which other baseline models do not perform. We provide the metrics for the abstaining predictions in Table \ref{table:baseline_comparisons_abstaining}. The models that do not perform abstaining predictions always predict the novel specimens incorrectly since these samples belongs to a  new class. The proposed ability to predict novel samples as \textit{"other"} may still require the model to have seen the given specimen in the \textit{"other"} class, which is not as effective as rejecting to make a prediction, which incorporates uncertainty in the model. In addition, as pointed out by Loi et al. \citep{loi2021towards}, predicting specimens as \textit{"other"} class often results in false predictions due to supervised models' common inability to learn patterns from a small number of samples. Our method novel ability to reject making a prediction yields promising results in identification of novel malware. Interestingly, around 22\% of the malware which we saw in the known set were also predicted as abstaining by the \textit{HNMFk Classifier}. This 22\% we referred as \emph{false-abstaining}, since the specimens here belongs to classes that we had labels for. Importantly, around 42\% of the novel malware (i.e. the malware which we did not see in the known set), are classified as abstaining. This 42\% is referred as \emph{true-abstaining} since our model did not have a reference label for these specimens in the known set. We also note that both true and false abstaining predictions would be caused by signatures or patterns extracted by NMFk being distinct from the labeled samples. Hence, it is possible that a detailed investigation and utilization of latent signatures can help to reveal characteristics that differ  given specimen from the known samples (similar to the prior work in latent mutational cancer signatures \cite{alexandrov2020repertoire}) and result in improvement of the abstaining predictions.

In Table \ref{table:baseline_comparisons_abstaining}, for completeness, we also provide F1 scores for each baseline that is calculated only of the specimens that HNMFk Classifier did make a predictions (i.e. it did not abstain). Notice that the F1 scores of our baselines increase, even surpass our model in some cases, when the  rejection to make predictions is not included in the score calculations. This result points out that the abstained samples are hard to correctly classify since our baselines yield lower scores when they are included (see the scored reported in Table \ref{table:baseline_comparisons_scores}). While the baselines falsely predicted the families for the harder specimens, HNMFk Classifier rejected to make a prediction and managed to maintain higher performance.

\begin{table*}[htb]
\caption{HNMFk Classifier is compared against the state-of-the-art supervised classifiers. The ability of the HNMFk to discover novel families is shown. \textbf{F1 - (Non-reject)} column shows the F1 scores for the specimens that HNMFk Classifier did make a prediction on. Not applicable (NA) used at the cells where the case does not apply to the given model. \textbf{Abstaining Seen} refers to false-abstaining predictions, samples that belong to known classes that were seen in the training set. Differently, \textbf{Abstaining Novel} shows the true-abstaining predictions, where the specimen belongs to a class that were not seen before.}
 \label{table:baseline_comparisons_abstaining}
\resizebox{\columnwidth}{!}{
\begin{tabular}{l|c|c|c}
\hline
\textbf{Model}                 & \textbf{Abstaining Seen (\%)}     & \textbf{Abstaining Novel (\%)} & \textbf{F1 - (Non-reject)}  \\ \hline
\text{HNMFk Classifier (semi-supervised)}       & 22.06           & \textbf{42.70}          &  0.80     \\
\text{HNMF2 Classifier (semi-supervised, ablation study)}       & \textbf{16.96}            & 34.16        & 0.77        \\
\text{XGBoost+SelfTrain (semi-supervised)}    & NA              & NA              & 0.81                    \\
\text{XGBoost (supervised)}                     & NA          & NA              & 0.80           \\   
\text{LightGBM (supervised, tuned on stratified subset)}        & NA           & NA              & 0.74            \\ 
\text{MLP (supervised)}                         & NA               & NA              & 0.79                    \\
\text{LightGBM+SelfTrain (semi-supervised)}    & NA              & NA              & 0.70                   \\
\hline
\end{tabular}
}
\vspace{-1.2em}
\end{table*}

We also apply our ablation study, where the number of cluster selection heuristic is turned off and rank-two factorization is used (i.e. $k=2$ at each node). In table \ref{table:baseline_comparisons_scores}, we can see that the \textit{HNMF2 Classifier} does perform better than our baseline models. However, the \textit{HNMFk Classifier} outperforms this method, which points out that carefully choosing the number of clusters improves the separability and the overall performance during prediction. \textit{HNMF2 Classifier} also reduces the percent of abstaining predictions, including the reduced percent of abstaining predictions on novel malware. We show additional results for ablation study on the automatic model selection below at Section \ref{sec:ablation_optimal_rank}.

Finally, note that in Table \ref{table:baseline_comparisons_scores} we have also included the tuning time comparison between the \textit{HNMFk Classifier} and the baseline models. The 5.77 days of tuning time listed for \textit{HNMFk Classifier} comes from our performance analysis on selecting the cluster uniformity threshold $t$ and understanding the effects of different hyper-parameter values of \textit{NMFk}. We selected $t=1$ for higher performance based on what we learned from the results of our experiments discussed in Section \ref{sec:cluster_quality}, and showed that the hyper-parameters of \textit{NMFk} has a minimal affect on the model's performance in Section \ref{sec:nmfk_hyperparameter_selection}. Note that the $k$ selection procedure of \textit{HNMFk Classifier}, which comes from the \textit{NMFk} algorithm, is not a hyper-parameter adjustment, but a model selection, which is integrated in the algorithm \cite{islam2021uncovering}; therefore, it is not included in the tuning time. Instead, it is reported as the  model training time. Our method takes about 8 days to complete running, which is significantly longer than our baseline models. In comparison to the traditional ML methods (in our case, the baseline models used in the experiments), our method is not a fast predictor. Instead, \textit{HNMFk Classifier} is a bulk-classification method. The aforementioned 8 days computation time is the total inference time for the \textit{HNMFk Classifier}. Therefore, our model is not suitable for real-time solutions, that is, for analysing of a single specimen at the time it comes in the system. Our method rather can be used for an accurate malware classification early in the labeling process. 

\subsection{Ablation Studies}
\label{sec:ablation_studies}

In our ablation studies we investigate the benefit of performing bulk classification and carefully choosing the number of clusters. To this end, during the first study we change the bulk classifier structure of our approach to form a more classical model, which we call the \textit{HNMFk Classical Classifier}. During the second study, we ablate the automatic model selection heuristic from our method. The small subset of the EMBER-2018 dataset, as described in Section \ref{sec:experiment_preperation}, is also used in our ablation studies in this section. As mentioned above, we use the top 10 malware families in the dataset with 1,000 specimens, each randomly sampled. Each experiment is run 10 times using a different random subset each time.  

\subsubsection{Bulk Classification}
\label{sec:bulk_classification}
\begin{figure}[htb]
\centering
\includegraphics[width=1\textwidth]{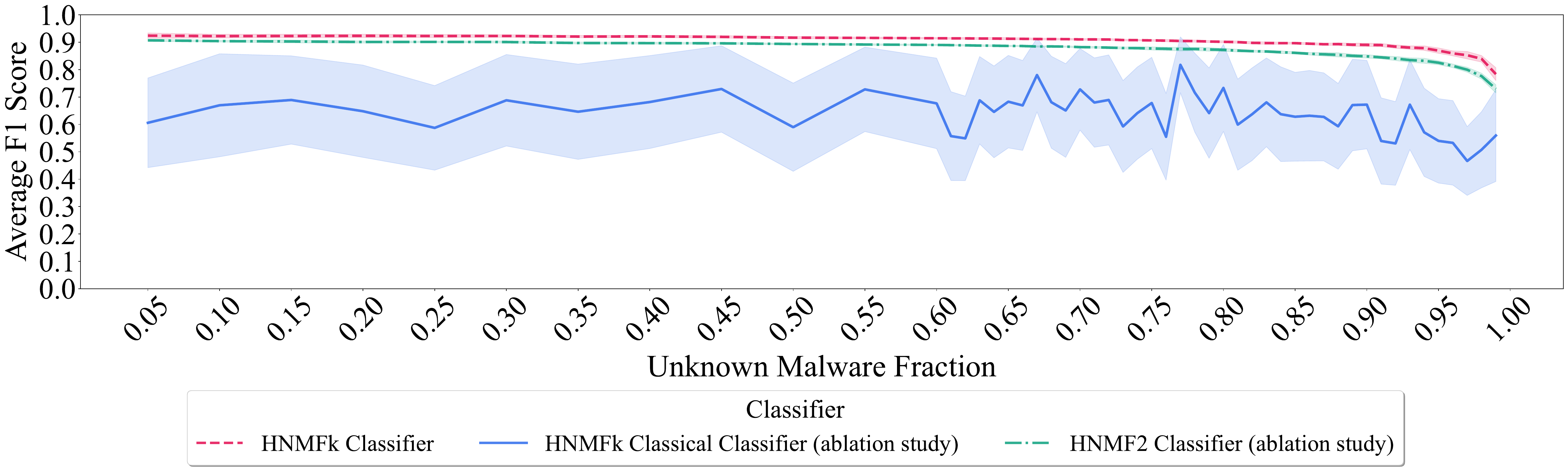}
\caption{The performance of the HNMFk Classifier is compared to the other variants of our method from the ablations studies, as the fraction of the unknown malware is changed. \label{fig:fraction_malware_fa_classifiers_ablation}}
\vspace{-1.2em}
\end{figure}

To show that there is a benefit to doing bulk classification for our methodology, we compare the performance of the \textit{HNMFk Classifier} to the \textit{HNMFk Classical Classifier}, a model that does not perform bulk classification. This model also uses the known samples to form the hierarchical graph. We then predict the unknown samples separately over the hierarchical graph by following the edges, and computing similarity scores at nodes. For each of the $n$ unknown malware samples, we obtain the cluster assignment by comparing the features $X_{i:}$ ($i$th sample) to the rows of the latent factor $\mat{H}$ using \textit{cosine-similarity} score:
\begin{equation}
    \label{eq:W_clustering2}
    \text{cluster}(i) = \underset{0\leq j \leq k^{opt}}{\operatorname{arg\,max}} \, (1-\text{cosine-distance}(\mat{H}_{ji}, \mat{X}_{i:}))
\end{equation}
We follow each sub-clusters, comparing the features vector for the $i$th sample to $\mat{H}$ at each step, until we reach a leaf where we predict the label of the specimen $i$ in a semi-supervised fashion. In Figure \ref{fig:fraction_malware_fa_classifiers_ablation} we compare the F1 scores obtained from our ablation studies to \textit{HNMFk Classifier} as the fraction of unknown samples change. From the figure, it can be seen that performing classification with \textit{HNMFk Classical Classifier} yields unstable results, and our method \textit{HNMFk Classifier} outperforms this model. This shows that bulk classification is beneficial in obtaining stable and accurate inference results.

\subsubsection{Determination of the Number of Clusters}
\label{sec:ablation_optimal_rank}
The \textit{HNMFk Classifier} utilizes the estimated number of components predicted by the \textit{NMFk} algorithm to achieve good separability of malware families. For the next ablation study, we look at the benefit of estimating $k$, or the number of clusters. During this study, we form another classifier named the \textit{HNMF2 Classifier}, based on the previous work of Gillis et al. \cite{gillis2014hierarchical}, which chooses $k=2$ at each node, i.e. separate the data into two clusters at each step, until each known sample falls in separate leaf nodes. 

In Figure \ref{fig:fraction_malware_fa_classifiers_ablation}, we also provide the results for the \textit{HNMF2 Classifier}. Choosing $k=2$ at each step performs almost as well as our approach. As also argued in \cite{gillis2014hierarchical}, this result points out the benefit of hierarchical setting. Even if we make a bad separation of the samples due to rank-two factorization, the hierarchical approach will fix the separations in the proceeding splits. However, although slightly, our model outperforms the \textit{HNMF2 Classifier}, which shows that choosing the number of components carefully using a heuristic is beneficial.

\section{Future Work}
\label{sec:future_work}

The \textit{HNMFk Classifier} has a significantly longer process time than the other ML methods which we used for comparison. The main cause for the increased computation time is the search of the number of clusters. \textit{NMFk} performs this search in a sequential manner, where each value of $k$ is tried, one after another. However, each rank $k$ factorization is independent from one another. Therefore, future work can consider parallelization of this task, or a distributed version of this task utilizing High-Performance Computing (HPC) environments~\cite{Boureima2022DistributedON,boureima2022distributed,bhattarai2023distributed,pyDNMFk1}.

Another future work includes the manual analysis of the specimens that fall in each cluster in the graph. It would be interesting to see which malware families are clustered together as we look at different nodes in the graph. This can help us understand if malware is clustered by type at first (such as botnet, backdoor, etc.), and then begin to separate into the families as we continue deeper in the graph. Future work can also include benign-ware as a class similar to \cite{8681127, loi2021towards, MOHAISEN2015251}.

We can also try to accelerate the computation time for clustering techniques via similarity-based approaches such as LZJD \cite{raff_lzjd_2017, raff_lzjd_digest} or BWMD \cite{Raff2020} by using the \textit{HNMFk Classifier} as a pre-processing step to obtain a hierarchical graph, where we then apply similarity comparisons only in the sub-trees instead of the entire dataset.    


\section{Conclusion}

In this paper, we introduced a novel semi-supervised classifier named the \textit{HNMFk Classifer}, that is capable of performing accurate bulk classification of thousands of malware families under extreme class imbalance conditions using the latent features extracted via \textit{NMFk}, which is used to perform automatic model selection, i.e. to estimate the number of clusters. Our method's ability to perform abstaining predictions allows it to maintain its accuracy when using a small amount of labeled data and when performing inference over novel malware families. In our experiments, we classified Windows malware using static malware analysis based features. \textit{HNMFk Classifer} is compared against the state-of-the-art baseline supervised and semi-supervised solutions, on which the prior work reported benchmarks, and surpassed their performance under the realistic experiment setting. Our new solution can be used to assist reverse engineers and malware analysts in the labeling process of malware families, outside the real-time environments. 

\begin{acks}
 This manuscript has been approved for unlimited release and has been assigned LA-UR-23-30350. We thank Nick Solovyev and Drew Barlow for helpful suggestions and edits. This research was partially funded by the Los Alamos National Laboratory (LANL) Laboratory Directed Research and Development (LDRD) grant 20190020DR and LANL Institutional Computing Program, supported by the U.S. Department of Energy National Nuclear Security Administration under Contract No. 89233218CNA000001.
\end{acks}

\bibliographystyle{ACM-Reference-Format}
\bibliography{references}

\end{document}